\documentclass[%
reprint,
superscriptaddress,
showpacs,preprintnumbers,
nofootinbib,
amsmath,assume,
aps,
prb
showkeys
]{revtex4-2}
\usepackage{graphicx}
\usepackage{dcolumn}
\usepackage{bm}

\usepackage{amssymb}
\usepackage{amsmath}
\usepackage{amsfonts}
\usepackage{xspace}
\usepackage{bm,bbm}
\usepackage{relsize}
\usepackage{siunitx}
\usepackage[usenames,dvipsnames]{color}
\usepackage{float}
\usepackage{hyperref}
\usepackage[toc,page]{appendix}
\usepackage{float}
\usepackage{array}
\newcolumntype{L}[1]{>{\raggedright\let\newline\\\arraybackslash\hspace{0pt}}m{#1}}
\newcolumntype{C}[1]{>{\centering\let\newline\\\arraybackslash\hspace{0pt}}m{#1}}
\newcolumntype{R}[1]{>{\raggedleft\let\newline\\\arraybackslash\hspace{0pt}}m{#1}}
\newcommand{\bigzero}{\mbox{\normalfont\Large\bfseries 0}}

\hypersetup{
    unicode=false,          
    pdftoolbar=true,        
    pdfmenubar=true,        
    pdffitwindow=false,     
    pdfstartview={FitH},    
    pdftitle={Optical bistability of zero-energy modes in Su-Schrieffer-Heeger lattice with Kerr nonlinearity},    
    pdfauthor={Alharbi, Wong, Gong, Leykam, Oh},     
    pdfsubject={},   
    pdfcreator={},   
    pdfproducer={}, 
    pdfkeywords={} {} {}, 
    pdfnewwindow=true,      
    colorlinks=true,       
    linkcolor=blue, 
    citecolor=blue,        
    filecolor=magenta,      
    urlcolor=blue,           
	breaklinks=true
}

\begin{document}


\title{Asymmetrical temporal dynamics of edge modes in Su-Schrieffer-Heeger lattice with Kerr nonlinearity} 

\author{Ghada H. Alharbi}
\affiliation{School of Physics and Astronomy, Cardiff University, Cardiff CF24 3AA, United Kingdom}
\affiliation{Department of Physics, University of Tabuk, Tabuk 741,  Kingdom of Saudi Arabia}
\author{Stephan Wong}
\affiliation{School of Physics and Astronomy, Cardiff University, Cardiff CF24 3AA, United Kingdom}
\affiliation{Center for Integrated Nanotechnologies, Sandia National Laboratories, Albuquerque, New Mexico 87185, USA}
\author{Yongkang Gong}
\affiliation{School of Physics and Astronomy, Cardiff University, Cardiff CF24 3AA, United Kingdom}
\author{Sang Soon Oh}
\email[Email:]{ohs2@cardiff.ac.uk}
\affiliation{School of Physics and Astronomy, Cardiff University, Cardiff CF24 3AA, United Kingdom}

\date{\today}

\begin{abstract}
Optical bistability and oscillating phases exist in a Sagnac interferometer and a single ring resonator made of $\chi^{(3)}$ nonlinear medium where the refractive indices are modulated by the light intensity due to the Kerr nonlinearity. An array of coupled nonlinear ring resonators behave similarly but with more complexity due to the presence of the additional couplings.
Here, we theoretically demonstrate the bifurcation of edge modes which leads to optical bistability in the Su-Schrieffer-Heeger lattice with the Kerr nonlinearity. Additionally, we demonstrate periodic and chaotic switching behaviors in an oscillating phase resulting from the coupling between the edge mode and bulk modes with different chiralities, i.e., clockwise and counter-clockwise circulations. 
\end{abstract}


\maketitle

\section{Introduction}
Asymmetrical states emerge when a system lose the balance and thus a symmetry between different components is broken. This can lead to bistability, where the system has two stable states for a single excitation. In some cases, asymmetrical dynamic states can emerge with periodic or chaotic oscillatory  behaviors. In photonic systems, optical bistability can appear when the light transmits through a cavity with a nonlinear medium leading to two different optical states, where one mode is dominant (switched on) and the other is quenched (switched off) \cite{Boyd2008}. 
For instance, the stable symmetry breaking has been studied for counter-propagating light beams in a Sagnac interferometer \cite{Asymmetrical1982}and micro-resonator with Kerr nonlinearity \cite{ DelBino2017a, Silver2018, Shim2016, Xu2021, coen_2024} and also for a singular direction of input light in a 1D chain of ring resonators \cite{Ghosh2024}. More interestingly, nonlinear optical ring resonators can present rich temporal dynamics with oscillatory behaviors, displaying various types of mode switching, such as chaotic, periodic, and self-switching dynamics \cite{Hill2020a}. These various dynamics are the result of the nonlinear interaction between the counter-propagating modes and they manifest the symmetry breaking, i.e., unequal intensities of the two counter-propagating modes.

Topological phases of matters are classified  depending on the symmetry and dimension, featuring topological defects and gapless modes in topological insulators and superconductors~\cite{Teo2010, Slager2019}. Further, three-dimensional topological band insulator can have dislocation-line modes~\cite{Slager2014}.
More recently, topologically protected modes have been widely studied in photonic systems due to their intriguing properties, such as unidirectional light propagation and robustness to defects and disorders~\cite{Ozawa2019, Khanikaev2013}. 
In particular, topological edge modes in a one-dimensional (1D) Su-Schrieffer-Heeger (SSH) configuration and two-dimensional (2D) photonic quantum spin-Hall or quantum valley-Hall structures have been employed in an array of coupled lasers, namely topological lasers \cite{Noh2018, Han2019, Harari2018b, Bandres2018, Gong2020, Wong2023}.
Moreover, nonlinear topological photonics has been studied in various platforms such as waveguide arrays \cite{Leykam2016, Zhou2017, Maczewsky2020, Ivanov2023}, microcavity polariton systems \cite{ Kartashov2017, Weifeng2019} and optical resonators \cite{Hadad2016, Dobrykh2018, Leykam2020,  Dobrykh2018, Roy2021, Ezawa2021, Ezawa2022, Wei2023}. 
The phase diagrams of a nonlinear SSH model and nonlinear breathing kagome model were drawn for the nonlinear parameters and coupling coefficients between sites \cite{Ezawa2021}. Also, the edge solitons have shown to be stable at any energy when the ratio between the weak and strong couplings falls below a critical value \cite{Ma2021}. Up until now, however, no research has demonstrated spontaneous symmetry breaking coming from the nonlinear response for the edge modes in photonic topological insulators.

In this paper, we theoretically show that we can observe asymmetrical temporal dynamics, including optical bistabilities and oscillation phases for edge modes in a nonlinear 1D SSH model. The system consists of an array of coupled ring resonators with the Kerr nonlinearity. 
Using the Lugiato-Lefever equation \cite{Lugiato1987} with additional nearest neighbor couplings, we demonstrate the optical bistability of the topological edge mode in the nonlinear SSH lattice. 
Finally, we use Poincar\'{e} section plots, composed of the maxima of the oscillating intensities, to display the oscillation phases featuring periodic and chaotic switching.

\section{Linear SSH model with two counter-propagating modes}
\begin{figure}
	\centering
	\includegraphics{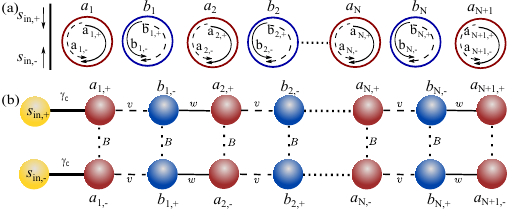}
	\caption{(a) A 1D array of coupled ring resonators with alternating gap sizes.  $s_{in,\pm}$ is the amplitude of input beams. (b) A schematic of the SSH model with resonators with Kerr nonlinearity. $B$ is the XPM strength, $v$ and $w$ are the nearest neighbor coupling coefficients between ring resonators, and $\gamma_c$ is the coupling coefficient between the waveguide and the first ring resonator.}
	\label{fig1}
\end{figure}

We start by considering a linear SSH chain which does not have any resonance frequency shift coming from a nonlinearity.
As shown in Fig.~\ref{fig1}(a), the one-dimensional chain has $(N+1)$ unit cells and every unit cell hosts two ring resonators; one on the sublattice A, and the other on the sublattice B. The $(N+1)$-th unit cell has only one ring resonator that belongs to the sublattice A, resulting in $M=2N+1$ ring resonators in total.  
In photonics, this SSH model can be implemented by alternating the gap size between the ring resonators, resulting in different intra- and inter-cell coupling coefficients $v$ and $w$ (Fig.~\ref{fig1}(b)). 
The coupled ring resonators are excited by two optical pumps with the same intensity, both of which are coupled into the first ring resonator but in the opposite directions exciting clockwise (CW) and counter-clockwise (CCW) modes, respectively. 
Then, the optical waves propagate back and forth through all the resonators via the couplings between the ring resonators.
To calculate the intensities of circulating optical waves at all ring resonators, we describe the time evolution of the field amplitudes $a_{n}(t)$ and $b_{n}(t)$ in the $n$-th unit cell (Fig.~\ref{fig1}(b)) for which we use the temporal coupled mode theory \cite{Haus1983, Suh2004}.  
Then, the coupled mode equations are written as:
\begin{eqnarray}
 \frac{da_{n,\pm}}{dt} &=&  i \Big(\omega_0  + i \gamma_n\Big)  a_{n,\pm} +  i v b_{n,\mp} +i w b_{n-1,\mp}  \nonumber \\ 
&&  + \delta_{n,1} \gamma_c s_{in} ,\nonumber \\ 
 \frac{db_{n,\pm}}{dt} &=&  i \Big(\omega_0  + i \gamma'_n\Big) b_{n,\pm}  + i v a_{n,\mp} + i  w a_{n+1,\mp} ,
\label{CMT}
\end{eqnarray}
where
\begin{eqnarray}
  \gamma_n &=& \gamma_0 + \delta_{n,1}\gamma_c ,\nonumber \\
  \gamma'_n &=& \gamma_0 .
\end{eqnarray}
Here, the subscript $\pm$ denotes the mode propagation directions, CW and CCW, respectively. 
$\delta_{n,1}$ is the Kronecker delta and $\omega_0$ is
the resonance frequency of the uncoupled ring resonators.
The two input beams with the same amplitude $s_{in}$, which is given as $\sqrt{I_s} e^{i\omega t}$ for pump intensity $I_s$, are coupled to the CW and CCW modes in the first ring ($a_{1,\pm}$) with the waveguide-to-ring coupling coefficient $\gamma_c$. Note that only $a_{n}$'s and $b_{n}$'s are time-dependent functions and we have omitted the symbol $(t)$ for brevity. 

To be more compact, we express the coupled mode equations (Eq.~(\ref{CMT})) in a matrix form by using the Hamiltonian $\mathbf{H}$ as
\begin{eqnarray}
\frac{d \mathbf{x} }{dt} = \mathbf{H} \mathbf{x} + \mathbf{S} 
\label{main_eq}
\end{eqnarray}
where 
\begin{equation*}
    \mathbf{x} = (a_{1,+} , b_{1,-},  a_{2,+},  b_{2,-}, \ldots, a_{1,-},  b_{1,+},  a_{2,-} , b_{2,+}, \ldots )^\mathsf{T}   . 
\end{equation*} 
Note that the Hamiltonian $\mathbf{H}$ can be differently defined after multiplying $i$ in both sides, which makes the equation look like the Schr\"{o}dinger equation and makes its eigenvalues correspond to the real parts of the frequencies. However, we have chosen this notation to make Eq.~(\ref{main_eq}) similar to Lugiato-Lefever equation which we will explain in the following section.  
Then, the Hamiltonian $\mathbf{H} $ can be split into two terms like:
\begin{equation}    
\mathbf{H} = \mathbf{H}_0 + \mathbf{H}_c
\label{linear_hamiltonian}
\end{equation}
where
\begin{equation}    
\mathbf{H}_0 = i (\omega_0 + i\gamma_0 ) \mathbb{I}_{2M}
\end{equation}
with $\mathbb{I}_{2M}$ the $(2M \times 2M)$ identity matrix.
The ring-to-ring coupling is expressed as:
\begin{equation}    	
\mathbf{H}_c = i \begin{pmatrix}
0 &v &0 & 0 &\cdots\\
v & 0 & w & 0 &\cdots\\
0 &  w & 0 & v  &\cdots\\
0 & 0  & v& 0& \cdots \\
\vdots& \vdots&\vdots& \vdots& \ddots
\end{pmatrix}.
\end{equation} 
Finally, the source term  $  \mathbf{S} $ is expressed as $ [s_{in,+}, 0, 0, \cdots,  s_{in,-}, 0, 0, \cdots ]^\mathsf{T} $. For the remainder of this paper, we assume the symmetric pumping by setting $s_{in,+} = s_{in,-}$.

The coupled mode equations (Eq.~(\ref{main_eq})) can be solved in both frequency and time domains. 
For example, in the frequency domain, by assuming $ \mathbf{x} = \mathbf{\tilde{x}} \exp(i \omega t)$ and $\mathbf{S} = 0$, we obtain an eigenvalue equation 
\begin{equation}   
i \omega \mathbf{\tilde{x}}  = \mathbf{H} \mathbf{\tilde{x}}
.
\label{frequency_domain}
\end{equation}
Solving the eigenvalue equation gives an frequency spectrum with so-called  zero-energy modes that are topologically protected and localized on one of the edges of the SSH chain with the smaller coupling coefficient among $v$ and $w$. In this work, we will use the term \emph{edge modes} because their frequencies deviate from the resonance frequency $\omega_0$ and thus they are not any more zero-energy modes for nonlinear cases. We call the rest of the modes \emph{bulk modes} as the mode fields are delocalized over the entire SSH lattice.

\section{Lugiato-Lefever equation for nonlinear SSH model}
Now we introduce the Kerr nonlinearity in the linear SSH model. 
In optics, the Kerr nonlinearity induces various nonlinear effects, for instance, self-phase modulation (SPM),  cross-phase modulation (XPM), four-wave mixing, and two-photon absorption \cite{Agrawal2019}. 
Here, we only consider the SPM and XPM for the counter-propating rotating modes in the ring resonators, both of which lead to a shift of the resonance frequencies of the CW or CCW modes. 
Although only the couplings due to the XPM are shown in Fig.~\ref{fig1}(b), the frequency shift $\Delta \omega$ is expressed by $(AI_{n,+} + BI_{n,-})$ where $I_{n,+}$ and $I_{n,-}$ are the intensities of the CW and CCW modes in the $n$-th ring resonator, respectively. $A$ and $B$ are  the SPM and XPM nonlinear coefficients, respectively.

For simplicity, we use the normalized Lugiato-Lefever equation to describe the field amplitudes in our nonlinear SSH model \cite{Lugiato1987}. Notably, the equation is equivalent to the one derived from the temporal coupled-mode theory (see Appendix \ref{appendix:LLeq} for more details). 
With the time-varying envelope amplitudes $\tilde{a}(t)$, $\tilde{b}(t)$, defined as $a(t)=\tilde{a}(t) e^{i\omega t}$, $b(t)=\tilde{b}(t) e^{i\omega t}$ respectively, the Lugiato-Lefever equation for a 1D SSH array of nonlinear ring resonators can be written as:
\begin{eqnarray}
\frac{d\tilde{a}_{n,\pm}}{d\Bar{t}} &=&  -\tilde{a}_{n,\pm}-i \eta \Delta \tilde{a}_{n,\pm} 
  + i \eta (A|\tilde{a}_{n,\pm}|^2 + B|\tilde{a}_{n,\mp}|^2)  \tilde{a}_{n,\pm} \nonumber \\ 
&& +i v \tilde{b}_{n\mp} + i w \tilde{b}_{n-1,\mp}  +\delta_{n,1}s_{in}, 
\nonumber \\ 
\frac{d\tilde{b}_{n,\pm}}{d\Bar{t}} &=&  -\tilde{b}_{n,\pm}-i \eta\Delta \tilde{b}_{n,\pm} + i \eta (A|\tilde{b}_{n,\pm}|^2 + B|\tilde{b}_{n,\mp}|^2) \tilde{b}_{n,\pm}   \nonumber \\ 
&& +i v \tilde{a}_{n\mp} + i w \tilde{a}_{n+1,\mp} ,
\label{nonlinear_LL_equation}
\end{eqnarray}
where $\Bar{t}=t\gamma_0$ is the dimensionless time.  The first term on the right-hand side represents damping, while the second term stands for detuning ($\Delta=(\omega-\omega_0)/\gamma_0$), which is the difference between the frequency of the continuous wave input beams and the resonance frequency of a single ring resonator. The third and fourth terms correspond to the SPM and XPM, respectively, with the normalized nonlinear coefficients $A$ and $B$, and $\eta = +1$ for a self-focusing medium or $\eta = -1$ for a self-defocusing medium. The terms with $v$ and $w$ refer to the intra- and inter- couplings between ring resonators as in the linear case.
Finally, we add a nonlinear term to Eq.~(\ref{linear_hamiltonian}) to have the Hamiltonian for our nonlinear SSH model: 

\begin{equation}    
\mathbf{H} = \mathbf{H}_0 + \mathbf{H}_c +  \mathbf{H}_{NL}  ,
\label{nonlinear_hamiltonian}
\end{equation}
where 
\begin{equation}    
\mathbf{H}_0 =  -(1 +i \eta \Delta )\mathbb{I}_{2 M},
\end{equation} and 

\begin{equation}    
\mathbf{H}_{NL} = i \eta  \mathbb{I}_{2M}\times
\mathbf{diag} \begin{pmatrix}
\mu_{1,a}^{cw},  &
\mu_{1,b}^{cw}, &
\hdots, &
\mu_{1,a}^{ccw},  &
\mu_{1,b}^{ccw}, &
\hdots
\end{pmatrix} ,
\end{equation}
where 
\begin{equation}
 \mu_{n,a}^{cw} = A|\tilde{a}_{n,+}|^2 + B|\tilde{a}_{n,-}|^2 , \nonumber   
 \end{equation}
  and 
  \begin{equation}
  \mu_{n,a}^{ccw} = A|\tilde{a}_{n,-}|^2 + B|\tilde{a}_{n,+}|^2 \nonumber .
  \end{equation}
 Similarly we can define $\mu_{n,b}^{cw}$ and $\mu_{n,b}^{ccw}$.
Because of the nonlinear terms, the time-dependent equation (Eq.(\ref{main_eq})) cannot be changed into an eigenvalue equation.
Instead, we can obtain the temporal evolution of the amplitudes $\tilde{a}_{n,\pm}$, $\tilde{b}_{n,\pm}$ by solving it numerically. 
In this work, we use the Runge-Kutta fourth order method to integrate Eq.(\ref{main_eq}) with respect to the time.

\begin{figure}
   \includegraphics{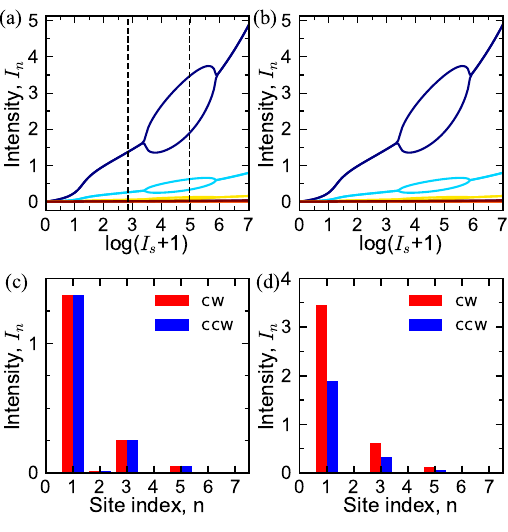}
	\caption{(a) Optical bistability for seven ring resonators ($M=7$) with detuning $\Delta=2.1$, $A=1$, $B=2.5$, $\gamma_c=1$, $v=3$ and $w=7$. (b) Optical bistability with the Kerr nonlinearity in the first ring resonator only. (c), (d) The distributions of intensity for the input intensities corresponding to the dashed vertical lines in (a).}\label{bistability}
\end{figure} 
\section{Optical bistability of topological edge modes}

\begin{figure*}
   \includegraphics{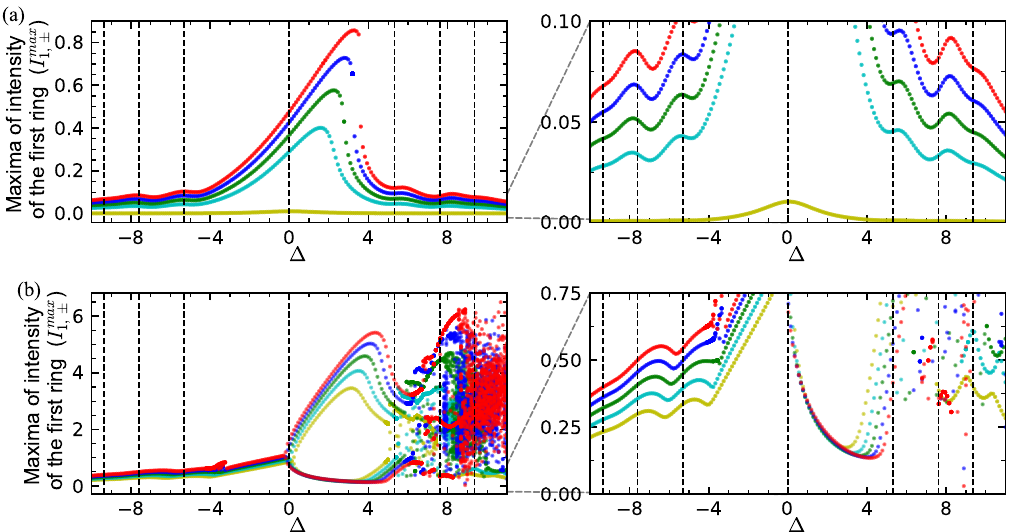}
	\caption{Poincar\'{e} section of the maxima of the CW and CCW intensity time series for the first ring resonator in 1D SSH lattice composed seven ring resonators with alternating coupling, $v=3$ and $w=7$, here $A=1$, $B=4$ and $\gamma_c=1$. (a) Maximum intensity curves for low input intensities $I_s=0.05$ (yellow), $2$ (cyan), $3$ (green), $4$ (blue), $5$ (red). (b) For high input intensities  $I_s=20$ (yellow), $25$ (cyan), $30$ (green), $35$ (blue), $40$ (red),  both optical bistability and oscillation regions appear. The vertical dashed lines indicate frequencies determined by the eigenvalue equation for the linear case (Eq.(\ref{frequency_domain})).} \label{compare_intensity}
\end{figure*} 
Optical bistability in a single ring resonator is a result of the XPM between two counter-propagating modes \cite{Kaplan1981}. 
This means that the Kerr nonlinearity leads to a shift in the resonance frequency of the two counter-propagating modes due to both SPM and XPM, and the coupling via XPM between them leads to spontaneous symmetry breaking above a certain threshold pump intensity \cite{DelBino2017a}. 
Here, we want to answer the question whether we can observe the optical bistability using an edge mode in a nonlinear SSH lattice model. 

To theoretically observe the optical bistability in the nonlinear SSH lattice, we consider a SSH array of seven ring resonators ($M=7$) with detuning of $2.1$, $A=1$, $B=2.5$ and $\gamma_c=1$ as and the alternating coupling coefficients ($v=3$ and $w=7$). In our simulations, we vary the pump intensity $I_s$ for a certain interval with random initial conditions. Indeed, as shown in Fig.~\ref{bistability}(a), we observe the optical bistability where the symmetry between CW and CCW modes is broken for the pump intensity range between $\log(I_s+1)=3.4$ and $\log(I_s+1)=5.9$. This means that one of the CW and CCW modes becomes stronger while the other mode is suppressed. As shown in Fig.~\ref{bistability} (c, d), the amplitude decreases gradually towards the bulk of the lattice system. 
Note that the amplitude is relatively large for odd sites only (sublattice A), and the intensity decreases exponentially along the right direction for both single stable (Fig.~\ref{bistability}(c)) and bistable cases (Fig.~\ref{bistability} (d)), which is the reminiscence of the zero-energy edge modes.

To explain the origin of the observed optical bistability, we hypothesize that the optical bistability comes from the symmetry breaking in the first ring only. 
First, optical bistability in a single ring resonator can occur when the pump intensity is above a certain threshold called a bifurcation point. This means the first ring will show the optical bistability first as we increase the pump intensity under an excitation close to the zero-energy frequency. Indeed, the detuning $\Delta = 1.85$ is smaller than the topological band gap ($2|v-w| = 8$) meaning the zero-energy edge mode is  dominantly excited even though the system is at off-resonance.
This is supported by the field intensity distribution in  Fig.~\ref{bistability}(d). 
As the intensities in the rest of rings are much smaller than the first ring (Fig.~\ref{bistability}(d)), only the first ring introduces bistability and 
the modes in the first ring couple to the other rings successively instead of having additional optical bistability from the rest of the rings. 
Second, to confirm this propagation of asymmetric intensities, we consider the Kerr effect only in the first ring resonator but keep all other parameters the same. This is equivalent to switching off the Kerr effect in the $2N$ ring resonators except the first ring resonator in our original setting.  
As shown in Fig.~\ref{bistability}(b), the intensity-intensity curve has almost identical shape as the original one except slight reduction in the range of $I_s$ and slight change in the difference between two counter-propagating mode intensities.

\begin{figure}
   \includegraphics{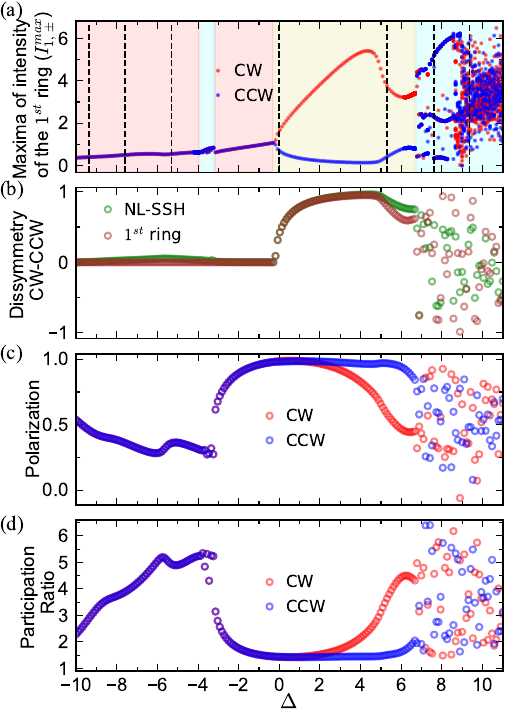}
	\caption{(a) Poincar\'{e} sections of the maxima of oscillating coupled intensity as a function of detuning for pump intensity $I_s=40$, for the first ring resonator from the 1D SSH array. The Poincar\'{e} sections for coupling coefficients ($v=3$, $w=7$) and XPM strength ($B$) of 4. The red shading indicates a symmetric case, the yellow indicates optical bistability, and the cyan indicates oscillations. The vertical dashed lines refer to resonance frequencies for the linear 1D SSH lattice. (b) Dissymmetry for CW and CCW intensity modes. (c) polarization in A and B sites. (d) Participation ratio for intensity.} \label{poincare_pr_c}
\end{figure} 

\begin{figure}   \includegraphics{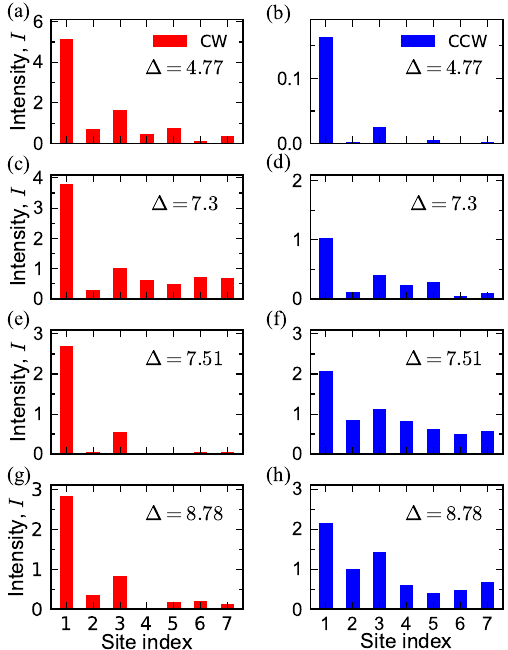}
	\caption{Snapshots of intensity distribution in the nonlinear 1D SSH lattice with $I_s=40$, $B=4$,  two figures in each row have the same detuning values for the CW modes and the CCW modes, respectively.  (a), (b) in the optical bistability regime, (c), (d) in the oscillation regime, and (e), (f)  in the oscillation switching regime.} \label{distribution_d}
\end{figure} 

\begin{figure}   
\includegraphics{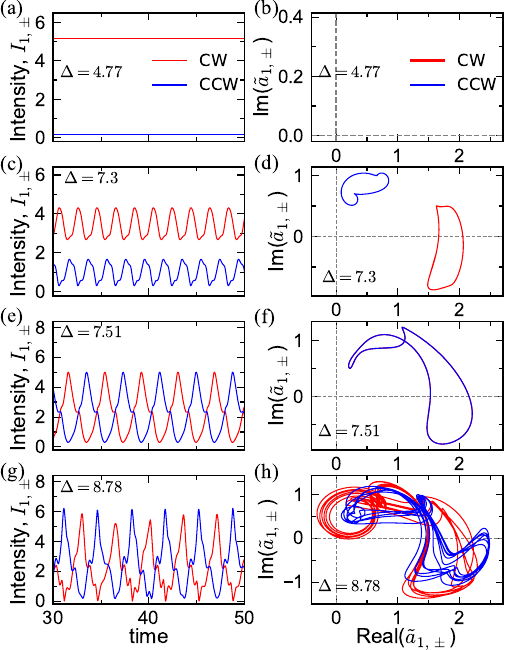}
	\caption{Time series of intensity and their phase space trajectories for $A=1$, $B=4$ and $I_s=40$ at different values of detuning, for the first ring resonators from 1 D SSH array of 7 rings. (a), (b) Optical bistability phase with $\Delta=4.77$. (c), (d) Oscillations without overlapping trajectories. (e), (f) Periodic switching for $\Delta=7.51$. (g), (h) Chaotic switching with $\Delta=8.78$. }\label{time_ev}
\end{figure} 


\section{Asymmetrical temporal dynamics}
Now, let us look at the temporal evolution of the optical intensities of a 1D SSH array that contains seven ring resonators.
To visualize oscillation and chaotic phases in our nonlinear system, we will use the Poincar\'{e} section 
obtained by plotting all the local maxima in a time series of oscillating intensities \cite{Hill2020a}. 
Since the intensity of each ring in the 1D SSH array follows the same pattern as the intensity of the first ring (see Fig.~\ref{odd_even} in Appendix \ref{appendix:poincare_SSH}), we plot the Poincar\'{e} sections for the first ring resonator only. 

As shown in Fig.~\ref{compare_intensity} (a)(b), the nonlinear SSH lattice exhibits both bistability and oscillation phases in the range of detuning corresponding to the edge mode and two bulk modes with positive detuning for the linear SSH lattice (denoted as the vertical dashed lines). 
Here, we set $A=1$ and $B=4$ and change the pump intensity from 0.05 to 20 denoted with different colors. 
As we can see in the zoomed view of the plots, these spectra show seven resonance modes; one edge mode with the largest intensity in the middle and six bulk modes on both sides of the edge mode having three on each side.  
For low pump intensities (Fig.~\ref{compare_intensity}(a)), the edge mode shifts to larger detuning values dramatically and its intensity increases significantly, whereas the bulk modes shift less and their intensities increase slightly. 
This is due to the localization of intensity at the first ring resonator. 
For high pump intensities (Fig.~\ref{compare_intensity}(b)), the CW and CCW modes for edge mode undergo an interaction between them via XPM, leading to an optical bistability. 
Remarkably, the high pump intensity leads to the interaction between the edge mode and the bulk mode near $\Delta = 6.8$ for $I_s=40$ as shown in Fig.~\ref{poincare_pr_c}(a), resulting in a series of oscillation phases occurring for both CW and CCW modes.
Note that CW and CCW intensities for bistable and oscillation phases can be exchanged any time because the chirality depends on the history of the dynamic system which is determined by the initial conditions in our case. 

To quantify the degree of the CW-CCW symmetry breaking (see Appendix \ref{appendix:Sublattice symmetry}), we introduce the dissymmetry parameter defined in a similar manner for a single microcavity in Ref. \cite{Cao2020} as:
\begin{equation}
        D = \frac{\sum_n I_{CW,n}-\sum_n I_{CCW,n}}{\sum_n I_{CW,n} + \sum_n I_{CCW,n}}.  
        \label{C}
\end{equation}
Then, a positive (negative) value of $D$ means the CW (CCW) mode is stronger than the other, corresponding to the optical bistability, and $D=0$ means the mode is CW-CCW symmetric. 
For the bistability regime, the nonlinear system shows relatively large dissymmetry close to 1 as shown in Fig.~\ref{poincare_pr_c}(b), meaning the CW-CCW symmetry is strongly broken. 
However, for the oscillations regime ($\Delta>6.9$), the dissymmetry has a random value between -1 and 1 and becomes very sensitive to the detuning. 
A similar trend can be observed in the dissymmetry calculated for the first ring only, reconfirming that the origin of the optical bistability is mainly due to the bifurcation of CW and CCW modes in the first ring.

To investigate the sublattice symmetry breaking (see Appendix \ref{appendix:Sublattice symmetry}), we calculate polarization ($P$) in terms of $\sum_n I_{A,n}$ (for sublattice A) and $\sum_n I_{B,n}$ (for sublattice B) for each direction using the following equation:
\begin{equation}
        P = \frac{\sum_n I_{A,n}-\sum_n I_{B,n}}{\sum_n I_{A,n} + \sum_n I_{B,n}}.  
        \label{polarization}
\end{equation}
Positive polarization values in Fig.~\ref{poincare_pr_c}(c) indicate that the modes for the bistability regime are localized at sublattice A; a value close to 1 confirms that the energy of the modes is concentrated at A sites. 
This is a reminiscence of topological zero-energy mode which would show a completely polarized distribution, i.e., $P=1$.

To characterize the localization of the intensity of dynamic modes in the whole SSH lattice, we calculate the participation ratio ($PR$), which is given by \cite{Longhi2018}:
\begin{equation}
        PR = \frac{(\sum_n I_n)^2}{\sum_n |I_n|^2}.
        \label{PR}
\end{equation}
A large value of $PR$ refers to delocalization.
At small values of detuning($\Delta \leq 1.5$),  as shown in Fig.~\ref{poincare_pr_c}(d), the nonlinear SSH lattice exhibits almost identical delocalization behavior for both CW and CCW modes. 
In contrast, for larger detuning ($\Delta \geq 1.5$), one of CW and CCW modes is more localized that the other.

To better understand the asymmetrical dynamic modes, we show the spatial distributions and temporal changes of the excited mode intensity for different detuning in Fig.~\ref{distribution_d} and Fig.~\ref{time_ev}. Here, we focus on the case of $I_s = 40$ and $B=4$ as an example of high pump intensity. 
For the optical bistability ($\Delta = 4.77$) shown in  Fig.~\ref{distribution_d}(a),(b), both CW and CCW modes have contrasting intensity values, whereas their profiles are similar to the zero-energy edge mode's profile in a linear SSH model with exponentially decaying non-zero odd-site intensities and zero even-site intensities. 
The deviations can be attributed to the off-resonance excitation and the interaction between CW and CCW modes via the nonlinear process (XPM). 
Note that the excited mode is stable as they have constant intensities and appear as two separate points in the phase space (Fig.~\ref{time_ev} (a),(b)). 
When we increase the detuning further to $\Delta = 7.3$, both CW and CCW mode profile deviates further away from the zero-energy edge mode but the CCW mode profile deviates less still having low intensities at even sites (Fig.~\ref{distribution_d}(c),(d)). Here, the largest value at the first site is related to the zero-energy edge mode and also due to the fact we are exciting the ring resonators from the waveguide on the left side. 
The dynamics for this detuning (Fig.~\ref{time_ev} (c)(d)) is periodically oscillatory, resulting in two distinct regions in the phase space meaning the CW mode intensity is always larger than the CCW mode intensity (the trajectory for CW is further away from the origin). 
For slightly larger detuning of $\Delta = 7.51$ (Fig.~\ref{time_ev} (e)(f)), the two trajectories are merged into one meaning that the intensities between the two modes alternate. In the phase space, they are located at different two points with the $\pi$ phase difference in the same trajectory. 
For a large detuning of $\Delta = 8.78$,  we see chaotic oscillations showing two separate trajectories covering a similar region in the phase space (Fig. ~\ref{time_ev}(g),(h)).

In contrast to the optical bistability coming from the coupling between two counter-propagating modes via nonlinearities (XPM and SPM), the emergence of the periodic and chaotic oscillations has its origin in the coupling between the edge mode and the bulk modes. 
The reasoning is as follows: 
Firstly, the resonance frequency shift of the edge mode when increasing the intensity is much larger than the ones for the bulk modes (Fig. ~\ref{poincare_pr_c}(a)), and
there are clear signatures of the edge modes, i.e., an exponentially decaying odd-site intensities and nearly zero even-site intensities (Fig.~\ref{distribution_d} (c-h) although the intensity distributions are getting close to the bulk modes. 
Thus, our numerical simulations confirm that the edge mode overlap with bulk mode due to the Kerr nonlinearity results in the periodic and chaotic oscillations.

\section{Conclusions}
In summary, we have numerically demonstrated the optical bistability and various types of oscillations in a 1D SSH model composed of ring resonators with Kerr nonlinearity. 
When the nonlinear terms are introduced in the Lugiato-Lefever equation, the first ring's CW and CCW mode intensities are symmetric until the pump intensity reaches a bifurcation point. Above the bifurcation point, the symmetry is spontaneously broken due to the splitting of the resonance frequencies of the two CW and CCW modes in the first ring resonator. 
For the high intensity regime, we have observed oscillating phases including periodic and chaotic oscillations.
We have classified the periodic oscillation phases into two different phases where the trajectories are separate or identical in the phase space of the mode intensities. 
This emergence of the oscillating phases can be attributed to the coupling between the edge mode and bulk mode due to the large shift of resonance peaks of the edge mode. 

We believe that our theoretical model and numerical results will provide valuable insight in understanding the complex dynamics in coupled nonlinear resonator systems with two chiral modes. 
In practice, the asymmetrical dynamic modes can be demonstrated experimentally using an optical ring resonator array, such as, a silicon microring resonator array \cite{Bogaerts2012}. 
In particular, the bifurcation states can be a building block for optical memory and switching devices allowing for storing and maintaining information by controlling the pump intensities depending on the direction of light circulation \cite{DelBino:21, Woodley2020}. Additionally, the various spatio-temporal dynamics could be applied to the stability analysis of coupled lasers.

\begin{acknowledgments}
We are grateful to Daniel Leykam for fruitful discussions. The work is part-funded by the European Regional Development Fund through the Welsh Government (80762-CU145 (East)).
\end{acknowledgments}

\appendix
\renewcommand{\thefigure}{A\arabic{figure}}
\setcounter{figure}{0}

\section{Derivation of the Lugiato-Lefever equation using the coupled mode theory}
\label{appendix:LLeq}
The coupled mode theory \cite{Haus1983} has been used to describe the field amplitude $a$ propagating in an optical ring resonator, which can be written as
\begin{equation}
    \frac{da}{dt}= i\omega_0 a - \gamma a + \gamma_c s.
    \label{cmt}
\end{equation}
Here $\omega_0$ refers to the resonance frequency, $\gamma $ and $\gamma_c$ are the damping and coupling with the source coefficient, respectively.
we can express field amplitude $a$ in terms of envelope amplitude $\tilde{a}$ as:
\begin{equation}
    a = \tilde{a} e^{i\omega t},
\end{equation}
by substituting in Eq.(\ref{cmt}) :
\begin{equation}
    \frac{d\tilde{a}}{dt} =  [i(\omega_0 - \omega )-\gamma]  \tilde{a}+ \gamma_c \tilde{s},
\end{equation}
where $-\tilde{\Delta}=\omega_0 - \omega$, then we can rewrite this equation in terms of detuning as :
\begin{equation}
    \frac{d\tilde{a}}{dt} =  
    (-\gamma - i\tilde{\Delta}) \tilde{a} + \gamma_c \tilde{s}.
\end{equation}
This equation is equivalent to the Lugiato-Lefever equation without nonlinearity terms; the terms in RHS correspond to damping, detuning, and source terms, respectively.

\section{Sublattice symmetry and CW-CCW symmetry}
\label{appendix:Sublattice symmetry}

In the linear regime, i.e., when $I_s \simeq 0$, our SSH model (Fig. \ref{fig1}(b)) has two distinctive symmetries called sublattice symmetry and CW-CCW symmetry. While both of them are called chiral symmetry in the literature, we use the terms sublattice symmetry and CW-CCW symmetry to avoid confusion.  
The Hamiltonian of the SSH model can be written in the following block matrix form:
\begin{equation} 
\mathbf{H} = 
\begin{pmatrix}

  \mathbf{H}_{cw}
&
 \bigzero \\

  \bigzero 
&
  \mathbf{H}_{ccw}

\end{pmatrix},
\end{equation}
where $\mathbf{H}_{cw}$ ($\mathbf{H}_{ccw}$) is the SSH Hamiltonian for CW (CCW) modes with the size of $M \times M$. They can be written as:

\begin{equation} 
\label{Hcw}
\mathbf{H}_{cw} = 
\begin{pmatrix}
 \mu_{1,a}^{cw} & v & 0 & 0 &  \hdots \\
  v &  \mu_{1,b}^{cw}  & w & 0 &  \hdots \\
  0 & w & \ddots  & \ddots  & \vdots \\
  \vdots & \vdots & \ddots & \ddots  & \ddots
\end{pmatrix} ,
\end{equation}

and

\begin{equation} 
\label{Hccw}
\mathbf{H}_{ccw} = 
\begin{pmatrix}
 \mu_{1,a}^{ccw} & v & 0 & 0 &  \hdots \\
  v &  \mu_{1,b}^{ccw}  & w & 0 &  \hdots \\
  0 & w & \ddots  & \ddots  & \vdots  \\
  \vdots & \vdots & \ddots & \ddots  & \ddots
\end{pmatrix},
\end{equation}
respectively. Here, the diagonal terms, corresponding to the onsite energy in a tight-binding description, represent the shift of resonance frequencies:
\begin{equation}
 \mu_{n,a}^{cw} = -\eta \Delta + A|\tilde{a}_{n,+}|^2 + B|\tilde{a}_{n,-}|^2 ,   
 \label{shift1}
\end{equation}
  and 
\begin{equation}
  \mu_{n,a}^{ccw} = -\eta \Delta + A|\tilde{a}_{n,-}|^2 + B|\tilde{a}_{n,+}|^2.
   \label{shift2}
\end{equation}
$\mu_{n,b}^{cw}$ and $\mu_{n,b}^{ccw}$ are defined similary.

First, let us check the sublattice symmetry of the nonlinear SSH Hamiltonian. 
The sublattice symmetry operator $\mathbf{\Gamma }$ for the subspace with the size of $M \times M$ is given by :
\begin{equation}
\mathbf{\Gamma }=
    \begin{pmatrix}
      -1       & 0      & 0      & \hdots \\ 
      0       & 1     & 0      & \hdots \\
      0       & 0      & -1      & \ddots \\ 
      \vdots  & \vdots & \ddots & \ddots
    \end{pmatrix}. 
\end{equation} 
If we apply $\mathbf{\Gamma }$ to $\mathbf{H}_{cw}$ ($\mathbf{H}_{ccw}$), we obtain

\begin{equation}
    \mathbf{\Gamma} \mathbf{H}_{cw} \mathbf{\Gamma}^\dagger = 
    \begin{pmatrix}
    \mu_{1,a}^{cw} & -v & 0 & 0 &  \hdots \\
  -v &  \mu_{1,b}^{cw}  & -w & 0 &  \hdots \\
  0 & -w & \ddots  & \ddots  & \vdots \\
  \vdots & \vdots & \ddots & \ddots  & \ddots    
    \end{pmatrix},
\end{equation}
and 
\begin{equation}
    \mathbf{\Gamma} \mathbf{H}_{ccw} \mathbf{\Gamma}^\dagger = 
    \begin{pmatrix}
    \mu_{1,a}^{ccw} & -v & 0 & 0 &  \hdots \\
    -v &  \mu_{1,b}^{ccw}  & -w & 0 &  \hdots \\
    0 & -w & \ddots  & \ddots  & \vdots  \\
  \vdots & \vdots & \ddots & \ddots  & \ddots    
    \end{pmatrix}.
\end{equation}

This means that the sublattice symmetry is broken in the strong nonlinear regime because the diagonal terms of $\mathbf{H}_{cw}$ and $\mathbf{H}_{ccw}$ do not vanish making  $\mathbf{\Gamma} \mathbf{H }\mathbf{\Gamma}^\dagger \neq -\mathbf{H}$.
In the weak nonlinearity regime, however, the diagonal terms approximately equal zero, satisfying $\mathbf{\Gamma} \mathbf{H }\mathbf{\Gamma}^\dagger = -\mathbf{H}$, the sublattice symmetry is recovered. Here, we assume the detuning is zero. 

Next, we check the CW-CCW symmetry. Let us define a block mirror symmetry operator matrix $\mathbf{P}$ as:
\begin{equation} 
\mathbf{P} = 
\begin{pmatrix}
  \bigzero
&
 \mathbb{I}_M \\
  \mathbb{I}_M
&
 \bigzero 
\end{pmatrix},
\end{equation}
where $\mathbb{I}_M$ is an identity matrix with size of $M \times M$. Then, we apply $\mathbf{P}$ to the full Hamiltonian as below.
\begin{equation}
    \mathbf{P} \mathbf{H }\mathbf{P}^{-1} = \begin{pmatrix}
   \mathbf{H}_{ccw}
&
 \bigzero \\
  \bigzero
&
  \mathbf{H}_{cw} 
\end{pmatrix}.
\end{equation}
From \ref{Hcw}, \ref{Hccw}, one can notice that $\mathbf{P} \mathbf{H }\mathbf{P}^{-1} = \mathbf{H }$ is satisfied only when $\tilde{a}_{n,-} = \tilde{a}_{n,+}$ which corresponds to the symmetrical dynamic modes in the weak nonlinear regime. For strong nonlinear regime, we obtain $\mathbf{P} \mathbf{H }\mathbf{P}^{-1} \neq \mathbf{H }$ giving rise to asymmetrical dynamics.

\section{Site-dependence of Poincar\'{e} sections for nonlinear SSH model}
\label{appendix:poincare_SSH}
\begin{figure*}
   \includegraphics{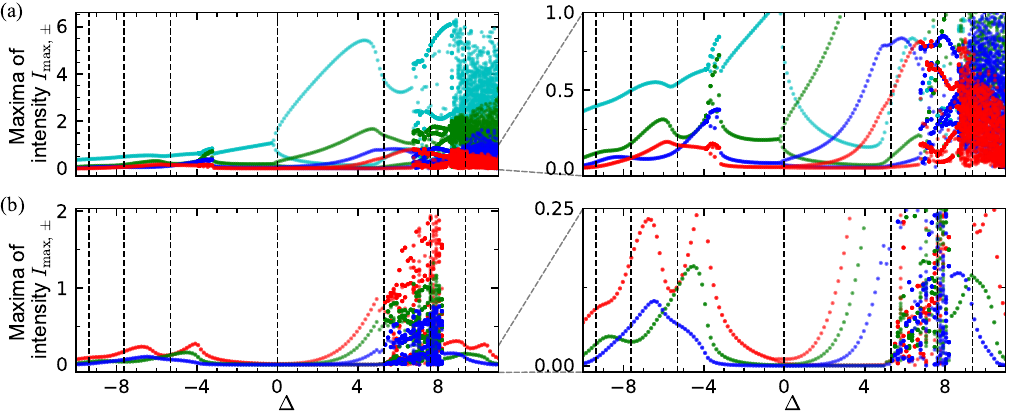}
	\caption{Poincar\'{e} sections of the maxima of oscillating coupled intensity as a function of detuning for a 1D SSH lattice ($M=7$) with the same parameters in Fig.~\ref{poincare_pr_c}(a). (a) For odd sites in the main text ($I_s=40, v=3, w=7, A=1, B=4, \gamma_c=1$) with the cyan, green, blue, and red colors corresponding to the 1st, 3rd, 5th, and 7th ring resonators respectively.  (b) For even sites with the red, green and blue colors corresponding to the 2nd, 4th, and 6th ring resonators, respectively.} \label{odd_even}
\end{figure*}

Figure~\ref{odd_even} (a), (b) displays Poincar\'{e} sections of maxima of oscillating in coupled intensities $I_{max,\pm}$ for odd (sublattice A) and even (sublattice B) sites in the 1D SSH lattice, respectively. The odd-site intensities of the CW and CCW modes follow the same pattern as the ones for the first ring resonator, while the even-site intensities follow the same patter as the one the second ring resonator. The symmetry is broken, i.e., the CW and CCW mode intensities are not equal for the optical bistability and oscillation phases.

\bibliography{ref}

\begin{thebibliography}{44}%
\makeatletter
\providecommand \@ifxundefined [1]{%
 \@ifx{#1\undefined}
}%
\providecommand \@ifnum [1]{%
 \ifnum #1\expandafter \@firstoftwo
 \else \expandafter \@secondoftwo
 \fi
}%
\providecommand \@ifx [1]{%
 \ifx #1\expandafter \@firstoftwo
 \else \expandafter \@secondoftwo
 \fi
}%
\providecommand \natexlab [1]{#1}%
\providecommand \enquote  [1]{``#1''}%
\providecommand \bibnamefont  [1]{#1}%
\providecommand \bibfnamefont [1]{#1}%
\providecommand \citenamefont [1]{#1}%
\providecommand \href@noop [0]{\@secondoftwo}%
\providecommand \href [0]{\begingroup \@sanitize@url \@href}%
\providecommand \@href[1]{\@@startlink{#1}\@@href}%
\providecommand \@@href[1]{\endgroup#1\@@endlink}%
\providecommand \@sanitize@url [0]{\catcode `\\12\catcode `\$12\catcode
  `\&12\catcode `\#12\catcode `\^12\catcode `\_12\catcode `\%12\relax}%
\providecommand \@@startlink[1]{}%
\providecommand \@@endlink[0]{}%
\providecommand \url  [0]{\begingroup\@sanitize@url \@url }%
\providecommand \@url [1]{\endgroup\@href {#1}{\urlprefix }}%
\providecommand \urlprefix  [0]{URL }%
\providecommand \Eprint [0]{\href }%
\providecommand \doibase [0]{https://doi.org/}%
\providecommand \selectlanguage [0]{\@gobble}%
\providecommand \bibinfo  [0]{\@secondoftwo}%
\providecommand \bibfield  [0]{\@secondoftwo}%
\providecommand \translation [1]{[#1]}%
\providecommand \BibitemOpen [0]{}%
\providecommand \bibitemStop [0]{}%
\providecommand \bibitemNoStop [0]{.\EOS\space}%
\providecommand \EOS [0]{\spacefactor3000\relax}%
\providecommand \BibitemShut  [1]{\csname bibitem#1\endcsname}%
\let\auto@bib@innerbib\@empty
\bibitem [{\citenamefont {Boyd}(2008)}]{Boyd2008}%
  \BibitemOpen
  \bibfield  {author} {\bibinfo {author} {\bibfnamefont {R.~W.}\ \bibnamefont
  {Boyd}},\ }\href@noop {} {\emph {\bibinfo {title} {{Nonlinear Optics}}}},\
  \bibinfo {edition} {3rd}\ ed.\ (\bibinfo  {publisher} {Academic Press, Inc},\
  \bibinfo {year} {2008})\BibitemShut {NoStop}%
\bibitem [{\citenamefont {Kaplan}\ and\ \citenamefont
  {Meystre}(1982)}]{Asymmetrical1982}%
  \BibitemOpen
  \bibfield  {author} {\bibinfo {author} {\bibfnamefont {A.}~\bibnamefont
  {Kaplan}}\ and\ \bibinfo {author} {\bibfnamefont {P.}~\bibnamefont
  {Meystre}},\ }\bibfield  {title} {\bibinfo {title} {{Directionally
  asymmetrical bistability in a symmetrically pumped nonlinear ring
  interferometer}},\ }\href {https://doi.org/10.1016/0030-4018(82)90267-X}
  {\bibfield  {journal} {\bibinfo  {journal} {Optics Communications}\ }\textbf
  {\bibinfo {volume} {40}},\ \bibinfo {pages} {229} (\bibinfo {year}
  {1982})}\BibitemShut {NoStop}%
\bibitem [{\citenamefont {{Del Bino}}\ \emph {et~al.}(2017)\citenamefont {{Del
  Bino}}, \citenamefont {Silver}, \citenamefont {Stebbings},\ and\
  \citenamefont {Del'Haye}}]{DelBino2017a}%
  \BibitemOpen
  \bibfield  {author} {\bibinfo {author} {\bibfnamefont {L.}~\bibnamefont {{Del
  Bino}}}, \bibinfo {author} {\bibfnamefont {J.~M.}\ \bibnamefont {Silver}},
  \bibinfo {author} {\bibfnamefont {S.~L.}\ \bibnamefont {Stebbings}},\ and\
  \bibinfo {author} {\bibfnamefont {P.}~\bibnamefont {Del'Haye}},\ }\bibfield
  {title} {\bibinfo {title} {{Symmetry Breaking of Counter-Propagating Light in
  a Nonlinear Resonator}},\ }\href {https://doi.org/10.1038/srep43142}
  {\bibfield  {journal} {\bibinfo  {journal} {Scientific Reports}\ }\textbf
  {\bibinfo {volume} {7}},\ \bibinfo {pages} {1} (\bibinfo {year}
  {2017})}\BibitemShut {NoStop}%
\bibitem [{\citenamefont {{Del Bino}}\ \emph {et~al.}(2018)\citenamefont {{Del
  Bino}}, \citenamefont {Silver}, \citenamefont {Woodley}, \citenamefont
  {Stebbings}, \citenamefont {Zhao},\ and\ \citenamefont
  {Del'Haye}}]{Silver2018}%
  \BibitemOpen
  \bibfield  {author} {\bibinfo {author} {\bibfnamefont {L.}~\bibnamefont {{Del
  Bino}}}, \bibinfo {author} {\bibfnamefont {J.~M.}\ \bibnamefont {Silver}},
  \bibinfo {author} {\bibfnamefont {M.~T.~M.}\ \bibnamefont {Woodley}},
  \bibinfo {author} {\bibfnamefont {S.~L.}\ \bibnamefont {Stebbings}}, \bibinfo
  {author} {\bibfnamefont {X.}~\bibnamefont {Zhao}},\ and\ \bibinfo {author}
  {\bibfnamefont {P.}~\bibnamefont {Del'Haye}},\ }\bibfield  {title} {\bibinfo
  {title} {{Microresonator isolators and circulators based on the intrinsic
  nonreciprocity of the Kerr effect}},\ }\href
  {https://doi.org/10.1364/OPTICA.5.000279} {\bibfield  {journal} {\bibinfo
  {journal} {Optica}\ }\textbf {\bibinfo {volume} {5}},\ \bibinfo {pages} {279}
  (\bibinfo {year} {2018})}\BibitemShut {NoStop}%
\bibitem [{\citenamefont {Shim}\ \emph {et~al.}(2016)\citenamefont {Shim},
  \citenamefont {Schlagheck}, \citenamefont {Hentschel},\ and\ \citenamefont
  {Wiersig}}]{Shim2016}%
  \BibitemOpen
  \bibfield  {author} {\bibinfo {author} {\bibfnamefont {J.-B.}\ \bibnamefont
  {Shim}}, \bibinfo {author} {\bibfnamefont {P.}~\bibnamefont {Schlagheck}},
  \bibinfo {author} {\bibfnamefont {M.}~\bibnamefont {Hentschel}},\ and\
  \bibinfo {author} {\bibfnamefont {J.}~\bibnamefont {Wiersig}},\ }\bibfield
  {title} {\bibinfo {title} {{Nonlinear dynamical tunneling of optical
  whispering gallery modes in the presence of a Kerr nonlinearity}},\ }\href
  {https://doi.org/10.1103/PhysRevA.94.053849} {\bibfield  {journal} {\bibinfo
  {journal} {Physical Review A}\ }\textbf {\bibinfo {volume} {94}},\ \bibinfo
  {pages} {053849} (\bibinfo {year} {2016})}\BibitemShut {NoStop}%
\bibitem [{\citenamefont {Xu}\ \emph {et~al.}(2021)\citenamefont {Xu},
  \citenamefont {Nielsen}, \citenamefont {Garbin}, \citenamefont {Hill},
  \citenamefont {Oppo}, \citenamefont {Fatome}, \citenamefont {Murdoch},
  \citenamefont {Coen},\ and\ \citenamefont {Erkintalo}}]{Xu2021}%
  \BibitemOpen
  \bibfield  {author} {\bibinfo {author} {\bibfnamefont {G.}~\bibnamefont
  {Xu}}, \bibinfo {author} {\bibfnamefont {A.~U.}\ \bibnamefont {Nielsen}},
  \bibinfo {author} {\bibfnamefont {B.}~\bibnamefont {Garbin}}, \bibinfo
  {author} {\bibfnamefont {L.}~\bibnamefont {Hill}}, \bibinfo {author}
  {\bibfnamefont {G.~L.}\ \bibnamefont {Oppo}}, \bibinfo {author}
  {\bibfnamefont {J.}~\bibnamefont {Fatome}}, \bibinfo {author} {\bibfnamefont
  {S.~G.}\ \bibnamefont {Murdoch}}, \bibinfo {author} {\bibfnamefont
  {S.}~\bibnamefont {Coen}},\ and\ \bibinfo {author} {\bibfnamefont
  {M.}~\bibnamefont {Erkintalo}},\ }\bibfield  {title} {\bibinfo {title}
  {{Spontaneous symmetry breaking of dissipative optical solitons in a
  two-component Kerr resonator}},\ }\href
  {https://doi.org/10.1038/s41467-021-24251-0} {\bibfield  {journal} {\bibinfo
  {journal} {Nature Communications}\ }\textbf {\bibinfo {volume} {12}},\
  \bibinfo {pages} {4023} (\bibinfo {year} {2021})}\BibitemShut {NoStop}%
\bibitem [{\citenamefont {Coen}\ \emph {et~al.}(2024)\citenamefont {Coen},
  \citenamefont {Garbin}, \citenamefont {Xu}, \citenamefont {Quinn},
  \citenamefont {Goldman}, \citenamefont {Oppo}, \citenamefont {Erkintalo},
  \citenamefont {Murdoch},\ and\ \citenamefont {Fatome}}]{coen_2024}%
  \BibitemOpen
  \bibfield  {author} {\bibinfo {author} {\bibfnamefont {S.}~\bibnamefont
  {Coen}}, \bibinfo {author} {\bibfnamefont {B.}~\bibnamefont {Garbin}},
  \bibinfo {author} {\bibfnamefont {G.}~\bibnamefont {Xu}}, \bibinfo {author}
  {\bibfnamefont {L.}~\bibnamefont {Quinn}}, \bibinfo {author} {\bibfnamefont
  {N.}~\bibnamefont {Goldman}}, \bibinfo {author} {\bibfnamefont {G.-L.}\
  \bibnamefont {Oppo}}, \bibinfo {author} {\bibfnamefont {M.}~\bibnamefont
  {Erkintalo}}, \bibinfo {author} {\bibfnamefont {S.~G.}\ \bibnamefont
  {Murdoch}},\ and\ \bibinfo {author} {\bibfnamefont {J.}~\bibnamefont
  {Fatome}},\ }\bibfield  {title} {\bibinfo {title} {Nonlinear topological
  symmetry protection in a dissipative system},\ }\href
  {https://doi.org/10.1038/s41467-023-44640-x} {\bibfield  {journal} {\bibinfo
  {journal} {Nature Communications}\ }\textbf {\bibinfo {volume} {15}},\
  \bibinfo {pages} {1398} (\bibinfo {year} {2024})}\BibitemShut {NoStop}%
\bibitem [{\citenamefont {Ghosh}\ \emph {et~al.}(2024)\citenamefont {Ghosh},
  \citenamefont {Pal}, \citenamefont {Hill}, \citenamefont {Campbell},
  \citenamefont {Bi}, \citenamefont {Zhang}, \citenamefont {Alabbadi},
  \citenamefont {Zhang}, \citenamefont {Oppo},\ and\ \citenamefont
  {Del'Haye}}]{Ghosh2024}%
  \BibitemOpen
  \bibfield  {author} {\bibinfo {author} {\bibfnamefont {A.}~\bibnamefont
  {Ghosh}}, \bibinfo {author} {\bibfnamefont {A.}~\bibnamefont {Pal}}, \bibinfo
  {author} {\bibfnamefont {L.}~\bibnamefont {Hill}}, \bibinfo {author}
  {\bibfnamefont {G.~N.}\ \bibnamefont {Campbell}}, \bibinfo {author}
  {\bibfnamefont {T.}~\bibnamefont {Bi}}, \bibinfo {author} {\bibfnamefont
  {Y.}~\bibnamefont {Zhang}}, \bibinfo {author} {\bibfnamefont
  {A.}~\bibnamefont {Alabbadi}}, \bibinfo {author} {\bibfnamefont
  {S.}~\bibnamefont {Zhang}}, \bibinfo {author} {\bibfnamefont {G.-L.}\
  \bibnamefont {Oppo}},\ and\ \bibinfo {author} {\bibfnamefont
  {P.}~\bibnamefont {Del'Haye}},\ }\bibfield  {title} {\bibinfo {title}
  {Controlled light distribution with coupled microresonator chains via kerr
  symmetry breaking},\ }\href@noop {} {\bibfield  {journal} {\bibinfo
  {journal} {arXiv}\ } (\bibinfo {year} {2024})},\ \Eprint
  {https://arxiv.org/abs/2402.10673} {2402.10673 [physics.optics]} \BibitemShut
  {NoStop}%
\bibitem [{\citenamefont {Hill}\ \emph {et~al.}(2020)\citenamefont {Hill},
  \citenamefont {Oppo}, \citenamefont {Woodley},\ and\ \citenamefont
  {Del'Haye}}]{Hill2020a}%
  \BibitemOpen
  \bibfield  {author} {\bibinfo {author} {\bibfnamefont {L.}~\bibnamefont
  {Hill}}, \bibinfo {author} {\bibfnamefont {G.-L.}\ \bibnamefont {Oppo}},
  \bibinfo {author} {\bibfnamefont {M.~T.~M.}\ \bibnamefont {Woodley}},\ and\
  \bibinfo {author} {\bibfnamefont {P.}~\bibnamefont {Del'Haye}},\ }\bibfield
  {title} {\bibinfo {title} {{Effects of self- and cross-phase modulation on
  the spontaneous symmetry breaking of light in ring resonators}},\ }\href
  {https://doi.org/10.1103/PhysRevA.101.013823} {\bibfield  {journal} {\bibinfo
   {journal} {Physical Review A}\ }\textbf {\bibinfo {volume} {101}},\ \bibinfo
  {pages} {013823} (\bibinfo {year} {2020})}\BibitemShut {NoStop}%
\bibitem [{\citenamefont {Teo}\ and\ \citenamefont {Kane}(2010)}]{Teo2010}%
  \BibitemOpen
  \bibfield  {author} {\bibinfo {author} {\bibfnamefont {J.~C.~Y.}\
  \bibnamefont {Teo}}\ and\ \bibinfo {author} {\bibfnamefont {C.~L.}\
  \bibnamefont {Kane}},\ }\bibfield  {title} {\bibinfo {title} {{Topological
  defects and gapless modes in insulators and superconductors}},\ }\href
  {https://doi.org/10.1103/PhysRevB.82.115120} {\bibfield  {journal} {\bibinfo
  {journal} {Physical Review B}\ }\textbf {\bibinfo {volume} {82}},\ \bibinfo
  {pages} {115120} (\bibinfo {year} {2010})}\BibitemShut {NoStop}%
\bibitem [{\citenamefont {Slager}(2019)}]{Slager2019}%
  \BibitemOpen
  \bibfield  {author} {\bibinfo {author} {\bibfnamefont {R.-J.}\ \bibnamefont
  {Slager}},\ }\bibfield  {title} {\bibinfo {title} {{The translational side of
  topological band insulators}},\ }\href
  {https://doi.org/10.1016/j.jpcs.2018.01.023} {\bibfield  {journal} {\bibinfo
  {journal} {Journal of Physics and Chemistry of Solids}\ }\textbf {\bibinfo
  {volume} {128}},\ \bibinfo {pages} {24} (\bibinfo {year} {2019})}\BibitemShut
  {NoStop}%
\bibitem [{\citenamefont {Slager}\ \emph {et~al.}(2014)\citenamefont {Slager},
  \citenamefont {Mesaros}, \citenamefont {Juri{\v{c}}i{\'{c}}},\ and\
  \citenamefont {Zaanen}}]{Slager2014}%
  \BibitemOpen
  \bibfield  {author} {\bibinfo {author} {\bibfnamefont {R.-J.}\ \bibnamefont
  {Slager}}, \bibinfo {author} {\bibfnamefont {A.}~\bibnamefont {Mesaros}},
  \bibinfo {author} {\bibfnamefont {V.}~\bibnamefont {Juri{\v{c}}i{\'{c}}}},\
  and\ \bibinfo {author} {\bibfnamefont {J.}~\bibnamefont {Zaanen}},\
  }\bibfield  {title} {\bibinfo {title} {{Interplay between electronic topology
  and crystal symmetry: Dislocation-line modes in topological band
  insulators}},\ }\href {https://doi.org/10.1103/PhysRevB.90.241403} {\bibfield
   {journal} {\bibinfo  {journal} {Physical Review B}\ }\textbf {\bibinfo
  {volume} {90}},\ \bibinfo {pages} {241403} (\bibinfo {year}
  {2014})}\BibitemShut {NoStop}%
\bibitem [{\citenamefont {Ozawa}\ \emph {et~al.}(2019)\citenamefont {Ozawa},
  \citenamefont {Price}, \citenamefont {Amo}, \citenamefont {Goldman},
  \citenamefont {Hafezi}, \citenamefont {Lu}, \citenamefont {Rechtsman},
  \citenamefont {Schuster}, \citenamefont {Simon}, \citenamefont {Zilberberg},\
  and\ \citenamefont {Carusotto}}]{Ozawa2019}%
  \BibitemOpen
  \bibfield  {author} {\bibinfo {author} {\bibfnamefont {T.}~\bibnamefont
  {Ozawa}}, \bibinfo {author} {\bibfnamefont {H.~M.}\ \bibnamefont {Price}},
  \bibinfo {author} {\bibfnamefont {A.}~\bibnamefont {Amo}}, \bibinfo {author}
  {\bibfnamefont {N.}~\bibnamefont {Goldman}}, \bibinfo {author} {\bibfnamefont
  {M.}~\bibnamefont {Hafezi}}, \bibinfo {author} {\bibfnamefont
  {L.}~\bibnamefont {Lu}}, \bibinfo {author} {\bibfnamefont {M.~C.}\
  \bibnamefont {Rechtsman}}, \bibinfo {author} {\bibfnamefont {D.}~\bibnamefont
  {Schuster}}, \bibinfo {author} {\bibfnamefont {J.}~\bibnamefont {Simon}},
  \bibinfo {author} {\bibfnamefont {O.}~\bibnamefont {Zilberberg}},\ and\
  \bibinfo {author} {\bibfnamefont {I.}~\bibnamefont {Carusotto}},\ }\bibfield
  {title} {\bibinfo {title} {{Topological photonics}},\ }\href
  {https://doi.org/10.1103/RevModPhys.91.015006} {\bibfield  {journal}
  {\bibinfo  {journal} {Reviews of Modern Physics}\ }\textbf {\bibinfo {volume}
  {91}},\ \bibinfo {pages} {015006} (\bibinfo {year} {2019})}\BibitemShut
  {NoStop}%
\bibitem [{\citenamefont {Khanikaev}\ \emph {et~al.}(2013)\citenamefont
  {Khanikaev}, \citenamefont {{Hossein Mousavi}}, \citenamefont {Tse},
  \citenamefont {Kargarian}, \citenamefont {MacDonald},\ and\ \citenamefont
  {Shvets}}]{Khanikaev2013}%
  \BibitemOpen
  \bibfield  {author} {\bibinfo {author} {\bibfnamefont {A.~B.}\ \bibnamefont
  {Khanikaev}}, \bibinfo {author} {\bibfnamefont {S.}~\bibnamefont {{Hossein
  Mousavi}}}, \bibinfo {author} {\bibfnamefont {W.-K.}\ \bibnamefont {Tse}},
  \bibinfo {author} {\bibfnamefont {M.}~\bibnamefont {Kargarian}}, \bibinfo
  {author} {\bibfnamefont {A.~H.}\ \bibnamefont {MacDonald}},\ and\ \bibinfo
  {author} {\bibfnamefont {G.}~\bibnamefont {Shvets}},\ }\bibfield  {title}
  {\bibinfo {title} {{Photonic topological insulators}},\ }\href
  {https://doi.org/10.1038/nmat3520} {\bibfield  {journal} {\bibinfo  {journal}
  {Nature Materials}\ }\textbf {\bibinfo {volume} {12}},\ \bibinfo {pages}
  {233} (\bibinfo {year} {2013})}\BibitemShut {NoStop}%
\bibitem [{\citenamefont {Noh}\ \emph {et~al.}(2018)\citenamefont {Noh},
  \citenamefont {Huang}, \citenamefont {Chen},\ and\ \citenamefont
  {Rechtsman}}]{Noh2018}%
  \BibitemOpen
  \bibfield  {author} {\bibinfo {author} {\bibfnamefont {J.}~\bibnamefont
  {Noh}}, \bibinfo {author} {\bibfnamefont {S.}~\bibnamefont {Huang}}, \bibinfo
  {author} {\bibfnamefont {K.~P.}\ \bibnamefont {Chen}},\ and\ \bibinfo
  {author} {\bibfnamefont {M.~C.}\ \bibnamefont {Rechtsman}},\ }\bibfield
  {title} {\bibinfo {title} {{Observation of Photonic Topological Valley Hall
  Edge States}},\ }\href {https://doi.org/10.1103/PhysRevLett.120.063902}
  {\bibfield  {journal} {\bibinfo  {journal} {Physical Review Letters}\
  }\textbf {\bibinfo {volume} {120}},\ \bibinfo {pages} {063902} (\bibinfo
  {year} {2018})}\BibitemShut {NoStop}%
\bibitem [{\citenamefont {Han}\ \emph {et~al.}(2019)\citenamefont {Han},
  \citenamefont {Lee}, \citenamefont {Callard}, \citenamefont {Seassal},\ and\
  \citenamefont {Jeon}}]{Han2019}%
  \BibitemOpen
  \bibfield  {author} {\bibinfo {author} {\bibfnamefont {C.}~\bibnamefont
  {Han}}, \bibinfo {author} {\bibfnamefont {M.}~\bibnamefont {Lee}}, \bibinfo
  {author} {\bibfnamefont {S.}~\bibnamefont {Callard}}, \bibinfo {author}
  {\bibfnamefont {C.}~\bibnamefont {Seassal}},\ and\ \bibinfo {author}
  {\bibfnamefont {H.}~\bibnamefont {Jeon}},\ }\bibfield  {title} {\bibinfo
  {title} {{Lasing at topological edge states in a photonic crystal L3
  nanocavity dimer array}},\ }\href {https://doi.org/10.1038/s41377-019-0149-7}
  {\bibfield  {journal} {\bibinfo  {journal} {Light: Science {\&}
  Applications}\ }\textbf {\bibinfo {volume} {8}},\ \bibinfo {pages} {40}
  (\bibinfo {year} {2019})}\BibitemShut {NoStop}%
\bibitem [{\citenamefont {Harari}\ \emph {et~al.}(2018)\citenamefont {Harari},
  \citenamefont {Bandres}, \citenamefont {Lumer}, \citenamefont {Rechtsman},
  \citenamefont {Chong}, \citenamefont {Khajavikhan}, \citenamefont
  {Christodoulides},\ and\ \citenamefont {Segev}}]{Harari2018b}%
  \BibitemOpen
  \bibfield  {author} {\bibinfo {author} {\bibfnamefont {G.}~\bibnamefont
  {Harari}}, \bibinfo {author} {\bibfnamefont {M.~A.}\ \bibnamefont {Bandres}},
  \bibinfo {author} {\bibfnamefont {Y.}~\bibnamefont {Lumer}}, \bibinfo
  {author} {\bibfnamefont {M.~C.}\ \bibnamefont {Rechtsman}}, \bibinfo {author}
  {\bibfnamefont {Y.~D.}\ \bibnamefont {Chong}}, \bibinfo {author}
  {\bibfnamefont {M.}~\bibnamefont {Khajavikhan}}, \bibinfo {author}
  {\bibfnamefont {D.~N.}\ \bibnamefont {Christodoulides}},\ and\ \bibinfo
  {author} {\bibfnamefont {M.}~\bibnamefont {Segev}},\ }\bibfield  {title}
  {\bibinfo {title} {{Topological insulator laser: Theory}},\ }\href
  {https://doi.org/10.1126/science.aar4003} {\bibfield  {journal} {\bibinfo
  {journal} {Science}\ }\textbf {\bibinfo {volume} {359}},\ \bibinfo {pages}
  {eaar4003} (\bibinfo {year} {2018})}\BibitemShut {NoStop}%
\bibitem [{\citenamefont {Bandres}\ \emph {et~al.}(2018)\citenamefont
  {Bandres}, \citenamefont {Wittek}, \citenamefont {Harari}, \citenamefont
  {Parto}, \citenamefont {Ren}, \citenamefont {Segev}, \citenamefont
  {Christodoulides},\ and\ \citenamefont {Khajavikhan}}]{Bandres2018}%
  \BibitemOpen
  \bibfield  {author} {\bibinfo {author} {\bibfnamefont {M.~A.}\ \bibnamefont
  {Bandres}}, \bibinfo {author} {\bibfnamefont {S.}~\bibnamefont {Wittek}},
  \bibinfo {author} {\bibfnamefont {G.}~\bibnamefont {Harari}}, \bibinfo
  {author} {\bibfnamefont {M.}~\bibnamefont {Parto}}, \bibinfo {author}
  {\bibfnamefont {J.}~\bibnamefont {Ren}}, \bibinfo {author} {\bibfnamefont
  {M.}~\bibnamefont {Segev}}, \bibinfo {author} {\bibfnamefont {D.~N.}\
  \bibnamefont {Christodoulides}},\ and\ \bibinfo {author} {\bibfnamefont
  {M.}~\bibnamefont {Khajavikhan}},\ }\bibfield  {title} {\bibinfo {title}
  {{Topological insulator laser: Experiments}},\ }\href
  {https://doi.org/10.1126/science.aar4005} {\bibfield  {journal} {\bibinfo
  {journal} {Science}\ }\textbf {\bibinfo {volume} {359}},\ \bibinfo {pages}
  {eaar4005} (\bibinfo {year} {2018})}\BibitemShut {NoStop}%
\bibitem [{\citenamefont {Gong}\ \emph {et~al.}(2020)\citenamefont {Gong},
  \citenamefont {Wong}, \citenamefont {Bennett}, \citenamefont {Huffaker},\
  and\ \citenamefont {Oh}}]{Gong2020}%
  \BibitemOpen
  \bibfield  {author} {\bibinfo {author} {\bibfnamefont {Y.}~\bibnamefont
  {Gong}}, \bibinfo {author} {\bibfnamefont {S.}~\bibnamefont {Wong}}, \bibinfo
  {author} {\bibfnamefont {A.~J.}\ \bibnamefont {Bennett}}, \bibinfo {author}
  {\bibfnamefont {D.~L.}\ \bibnamefont {Huffaker}},\ and\ \bibinfo {author}
  {\bibfnamefont {S.~S.}\ \bibnamefont {Oh}},\ }\bibfield  {title} {\bibinfo
  {title} {{Topological Insulator Laser Using Valley-Hall Photonic Crystals}},\
  }\href {https://doi.org/10.1021/acsphotonics.0c00521} {\bibfield  {journal}
  {\bibinfo  {journal} {ACS Photonics}\ }\textbf {\bibinfo {volume} {7}},\
  \bibinfo {pages} {2089} (\bibinfo {year} {2020})}\BibitemShut {NoStop}%
\bibitem [{\citenamefont {Wong}\ \emph {et~al.}(2023)\citenamefont {Wong},
  \citenamefont {Olthaus}, \citenamefont {Bracht}, \citenamefont {Reiter},\
  and\ \citenamefont {Oh}}]{Wong2023}%
  \BibitemOpen
  \bibfield  {author} {\bibinfo {author} {\bibfnamefont {S.}~\bibnamefont
  {Wong}}, \bibinfo {author} {\bibfnamefont {J.}~\bibnamefont {Olthaus}},
  \bibinfo {author} {\bibfnamefont {T.~K.}\ \bibnamefont {Bracht}}, \bibinfo
  {author} {\bibfnamefont {D.~E.}\ \bibnamefont {Reiter}},\ and\ \bibinfo
  {author} {\bibfnamefont {S.~S.}\ \bibnamefont {Oh}},\ }\bibfield  {title}
  {\bibinfo {title} {{A machine learning approach to drawing phase diagrams of
  topological lasing modes}},\ }\href
  {https://doi.org/10.1038/s42005-023-01230-z} {\bibfield  {journal} {\bibinfo
  {journal} {Communications Physics}\ }\textbf {\bibinfo {volume} {6}},\
  \bibinfo {pages} {104} (\bibinfo {year} {2023})}\BibitemShut {NoStop}%
\bibitem [{\citenamefont {Leykam}\ and\ \citenamefont
  {Chong}(2016)}]{Leykam2016}%
  \BibitemOpen
  \bibfield  {author} {\bibinfo {author} {\bibfnamefont {D.}~\bibnamefont
  {Leykam}}\ and\ \bibinfo {author} {\bibfnamefont {Y.~D.}\ \bibnamefont
  {Chong}},\ }\bibfield  {title} {\bibinfo {title} {{Edge Solitons in
  Nonlinear-Photonic Topological Insulators}},\ }\href
  {https://doi.org/10.1103/PhysRevLett.117.143901} {\bibfield  {journal}
  {\bibinfo  {journal} {Physical Review Letters}\ }\textbf {\bibinfo {volume}
  {117}},\ \bibinfo {pages} {143901} (\bibinfo {year} {2016})}\BibitemShut
  {NoStop}%
\bibitem [{\citenamefont {Zhou}\ \emph {et~al.}(2017)\citenamefont {Zhou},
  \citenamefont {Wang}, \citenamefont {Leykam},\ and\ \citenamefont
  {Chong}}]{Zhou2017}%
  \BibitemOpen
  \bibfield  {author} {\bibinfo {author} {\bibfnamefont {X.}~\bibnamefont
  {Zhou}}, \bibinfo {author} {\bibfnamefont {Y.}~\bibnamefont {Wang}}, \bibinfo
  {author} {\bibfnamefont {D.}~\bibnamefont {Leykam}},\ and\ \bibinfo {author}
  {\bibfnamefont {Y.}~\bibnamefont {Chong}},\ }\bibfield  {title} {\bibinfo
  {title} {{Optical isolation with nonlinear topological photonics}},\ }\href
  {https://doi.org/10.1088/1367-2630/aa7cb5} {\bibfield  {journal} {\bibinfo
  {journal} {New Journal of Physics}\ }\textbf {\bibinfo {volume} {19}},\
  \bibinfo {pages} {095002} (\bibinfo {year} {2017})}\BibitemShut {NoStop}%
\bibitem [{\citenamefont {Maczewsky}\ \emph {et~al.}(2020)\citenamefont
  {Maczewsky}, \citenamefont {Heinrich}, \citenamefont {Kremer}, \citenamefont
  {Ivanov}, \citenamefont {Ehrhardt}, \citenamefont {Martinez}, \citenamefont
  {Kartashov}, \citenamefont {Konotop}, \citenamefont {Torner}, \citenamefont
  {Bauer},\ and\ \citenamefont {Szameit}}]{Maczewsky2020}%
  \BibitemOpen
  \bibfield  {author} {\bibinfo {author} {\bibfnamefont {L.~J.}\ \bibnamefont
  {Maczewsky}}, \bibinfo {author} {\bibfnamefont {M.}~\bibnamefont {Heinrich}},
  \bibinfo {author} {\bibfnamefont {M.}~\bibnamefont {Kremer}}, \bibinfo
  {author} {\bibfnamefont {S.~K.}\ \bibnamefont {Ivanov}}, \bibinfo {author}
  {\bibfnamefont {M.}~\bibnamefont {Ehrhardt}}, \bibinfo {author}
  {\bibfnamefont {F.}~\bibnamefont {Martinez}}, \bibinfo {author}
  {\bibfnamefont {Y.~V.}\ \bibnamefont {Kartashov}}, \bibinfo {author}
  {\bibfnamefont {V.~V.}\ \bibnamefont {Konotop}}, \bibinfo {author}
  {\bibfnamefont {L.}~\bibnamefont {Torner}}, \bibinfo {author} {\bibfnamefont
  {D.}~\bibnamefont {Bauer}},\ and\ \bibinfo {author} {\bibfnamefont
  {A.}~\bibnamefont {Szameit}},\ }\bibfield  {title} {\bibinfo {title}
  {{Nonlinearity-induced photonic topological insulator}},\ }\href
  {https://doi.org/10.1126/science.abd2033} {\bibfield  {journal} {\bibinfo
  {journal} {Science}\ }\textbf {\bibinfo {volume} {370}},\ \bibinfo {pages}
  {701} (\bibinfo {year} {2020})}\BibitemShut {NoStop}%
\bibitem [{\citenamefont {Ivanov}\ \emph {et~al.}(2023)\citenamefont {Ivanov},
  \citenamefont {Kartashov},\ and\ \citenamefont {Torner}}]{Ivanov2023}%
  \BibitemOpen
  \bibfield  {author} {\bibinfo {author} {\bibfnamefont {S.~K.}\ \bibnamefont
  {Ivanov}}, \bibinfo {author} {\bibfnamefont {Y.~V.}\ \bibnamefont
  {Kartashov}},\ and\ \bibinfo {author} {\bibfnamefont {L.}~\bibnamefont
  {Torner}},\ }\bibfield  {title} {\bibinfo {title} {{Light bullets in
  Su-Schrieffer-Heeger photonic topological insulators}},\ }\href
  {https://doi.org/10.1103/PhysRevA.107.033514} {\bibfield  {journal} {\bibinfo
   {journal} {Physical Review A}\ }\textbf {\bibinfo {volume} {107}},\ \bibinfo
  {pages} {033514} (\bibinfo {year} {2023})}\BibitemShut {NoStop}%
\bibitem [{\citenamefont {Kartashov}\ and\ \citenamefont
  {Skryabin}(2017)}]{Kartashov2017}%
  \BibitemOpen
  \bibfield  {author} {\bibinfo {author} {\bibfnamefont {Y.~V.}\ \bibnamefont
  {Kartashov}}\ and\ \bibinfo {author} {\bibfnamefont {D.~V.}\ \bibnamefont
  {Skryabin}},\ }\bibfield  {title} {\bibinfo {title} {{Bistable Topological
  Insulator with Exciton-Polaritons}},\ }\href
  {https://doi.org/10.1103/Phys.Rev.Lett.119.253904} {\bibfield  {journal}
  {\bibinfo  {journal} {Physical Review Letters}\ }\textbf {\bibinfo {volume}
  {119}},\ \bibinfo {pages} {253904} (\bibinfo {year} {2017})}\BibitemShut
  {NoStop}%
\bibitem [{\citenamefont {Zhang}\ \emph {et~al.}(2019)\citenamefont {Zhang},
  \citenamefont {Chen}, \citenamefont {Kartashov}, \citenamefont {Skryabin},\
  and\ \citenamefont {Ye}}]{Weifeng2019}%
  \BibitemOpen
  \bibfield  {author} {\bibinfo {author} {\bibfnamefont {W.}~\bibnamefont
  {Zhang}}, \bibinfo {author} {\bibfnamefont {X.}~\bibnamefont {Chen}},
  \bibinfo {author} {\bibfnamefont {Y.~V.}\ \bibnamefont {Kartashov}}, \bibinfo
  {author} {\bibfnamefont {D.~V.}\ \bibnamefont {Skryabin}},\ and\ \bibinfo
  {author} {\bibfnamefont {F.}~\bibnamefont {Ye}},\ }\bibfield  {title}
  {\bibinfo {title} {Finite‐dimensional bistable topological insulators: From
  small to large},\ }\href@noop {} {\bibfield  {journal} {\bibinfo  {journal}
  {Laser Photon. Rev.}\ }\textbf {\bibinfo {volume} {13}},\ \bibinfo {pages}
  {1900198} (\bibinfo {year} {2019})}\BibitemShut {NoStop}%
\bibitem [{\citenamefont {Hadad}\ \emph {et~al.}(2016)\citenamefont {Hadad},
  \citenamefont {Khanikaev},\ and\ \citenamefont {Al{\`{u}}}}]{Hadad2016}%
  \BibitemOpen
  \bibfield  {author} {\bibinfo {author} {\bibfnamefont {Y.}~\bibnamefont
  {Hadad}}, \bibinfo {author} {\bibfnamefont {A.~B.}\ \bibnamefont
  {Khanikaev}},\ and\ \bibinfo {author} {\bibfnamefont {A.}~\bibnamefont
  {Al{\`{u}}}},\ }\bibfield  {title} {\bibinfo {title} {{Self-induced
  topological transitions and edge states supported by nonlinear staggered
  potentials}},\ }\href {https://doi.org/10.1103/PhysRevB.93.155112} {\bibfield
   {journal} {\bibinfo  {journal} {Physical Review B}\ }\textbf {\bibinfo
  {volume} {93}},\ \bibinfo {pages} {155112} (\bibinfo {year}
  {2016})}\BibitemShut {NoStop}%
\bibitem [{\citenamefont {Dobrykh}\ \emph {et~al.}(2018)\citenamefont
  {Dobrykh}, \citenamefont {Yulin}, \citenamefont {Slobozhanyuk}, \citenamefont
  {Poddubny},\ and\ \citenamefont {Kivshar}}]{Dobrykh2018}%
  \BibitemOpen
  \bibfield  {author} {\bibinfo {author} {\bibfnamefont {D.~A.}\ \bibnamefont
  {Dobrykh}}, \bibinfo {author} {\bibfnamefont {A.~V.}\ \bibnamefont {Yulin}},
  \bibinfo {author} {\bibfnamefont {A.~P.}\ \bibnamefont {Slobozhanyuk}},
  \bibinfo {author} {\bibfnamefont {A.~N.}\ \bibnamefont {Poddubny}},\ and\
  \bibinfo {author} {\bibfnamefont {Y.~S.}\ \bibnamefont {Kivshar}},\
  }\bibfield  {title} {\bibinfo {title} {{Nonlinear Control of Electromagnetic
  Topological Edge States}},\ }\href
  {https://doi.org/10.1103/PhysRevLett.121.163901} {\bibfield  {journal}
  {\bibinfo  {journal} {Physical Review Letters}\ }\textbf {\bibinfo {volume}
  {121}},\ \bibinfo {pages} {163901} (\bibinfo {year} {2018})}\BibitemShut
  {NoStop}%
\bibitem [{\citenamefont {Leykam}\ and\ \citenamefont
  {Yuan}(2020)}]{Leykam2020}%
  \BibitemOpen
  \bibfield  {author} {\bibinfo {author} {\bibfnamefont {D.}~\bibnamefont
  {Leykam}}\ and\ \bibinfo {author} {\bibfnamefont {L.}~\bibnamefont {Yuan}},\
  }\bibfield  {title} {\bibinfo {title} {{Topological phases in ring
  resonators: recent progress and future prospects}},\ }\href
  {https://doi.org/10.1515/nanoph-2020-0415} {\bibfield  {journal} {\bibinfo
  {journal} {Nanophotonics}\ }\textbf {\bibinfo {volume} {9}},\ \bibinfo
  {pages} {4473} (\bibinfo {year} {2020})}\BibitemShut {NoStop}%
\bibitem [{\citenamefont {Roy}\ \emph {et~al.}(2022)\citenamefont {Roy},
  \citenamefont {Parto}, \citenamefont {Nehra}, \citenamefont {Leefmans},\ and\
  \citenamefont {Marandi}}]{Roy2021}%
  \BibitemOpen
  \bibfield  {author} {\bibinfo {author} {\bibfnamefont {A.}~\bibnamefont
  {Roy}}, \bibinfo {author} {\bibfnamefont {M.}~\bibnamefont {Parto}}, \bibinfo
  {author} {\bibfnamefont {R.}~\bibnamefont {Nehra}}, \bibinfo {author}
  {\bibfnamefont {C.}~\bibnamefont {Leefmans}},\ and\ \bibinfo {author}
  {\bibfnamefont {A.}~\bibnamefont {Marandi}},\ }\bibfield  {title} {\bibinfo
  {title} {{Topological optical parametric oscillation}},\ }\href
  {https://doi.org/10.1515/nanoph-2021-0765} {\bibfield  {journal} {\bibinfo
  {journal} {Nanophotonics}\ }\textbf {\bibinfo {volume} {11}},\ \bibinfo
  {pages} {1611} (\bibinfo {year} {2022})}\BibitemShut {NoStop}%
\bibitem [{\citenamefont {Ezawa}(2021)}]{Ezawa2021}%
  \BibitemOpen
  \bibfield  {author} {\bibinfo {author} {\bibfnamefont {M.}~\bibnamefont
  {Ezawa}},\ }\bibfield  {title} {\bibinfo {title} {{Nonlinearity-induced
  transition in the nonlinear Su-Schrieffer-Heeger model and a nonlinear
  higher-order topological system}},\ }\href
  {https://doi.org/10.1103/PhysRevB.104.235420} {\bibfield  {journal} {\bibinfo
   {journal} {Physical Review B}\ }\textbf {\bibinfo {volume} {104}},\ \bibinfo
  {pages} {235420} (\bibinfo {year} {2021})}\BibitemShut {NoStop}%
\bibitem [{\citenamefont {Ezawa}(2022)}]{Ezawa2022}%
  \BibitemOpen
  \bibfield  {author} {\bibinfo {author} {\bibfnamefont {M.}~\bibnamefont
  {Ezawa}},\ }\bibfield  {title} {\bibinfo {title} {{Nonlinear non-Hermitian
  higher-order topological laser}},\ }\href
  {https://doi.org/10.1103/PhysRevResearch.4.013195} {\bibfield  {journal}
  {\bibinfo  {journal} {Physical Review Research}\ }\textbf {\bibinfo {volume}
  {4}},\ \bibinfo {pages} {013195} (\bibinfo {year} {2022})}\BibitemShut
  {NoStop}%
\bibitem [{\citenamefont {Wei}\ \emph {et~al.}(2023)\citenamefont {Wei},
  \citenamefont {Liao}, \citenamefont {Wang}, \citenamefont {Zhu},
  \citenamefont {Xu},\ and\ \citenamefont {Yang}}]{Wei2023}%
  \BibitemOpen
  \bibfield  {author} {\bibinfo {author} {\bibfnamefont {M.-S.}\ \bibnamefont
  {Wei}}, \bibinfo {author} {\bibfnamefont {M.-J.}\ \bibnamefont {Liao}},
  \bibinfo {author} {\bibfnamefont {C.}~\bibnamefont {Wang}}, \bibinfo {author}
  {\bibfnamefont {C.}~\bibnamefont {Zhu}}, \bibinfo {author} {\bibfnamefont
  {J.}~\bibnamefont {Xu}},\ and\ \bibinfo {author} {\bibfnamefont
  {Y.}~\bibnamefont {Yang}},\ }\bibfield  {title} {\bibinfo {title} {{Phase
  transition and dynamics of qubits in coupled-cavity arrays with nonlinear
  topological photonics}},\ }\href {https://doi.org/10.1016/j.rinp.2023.106232}
  {\bibfield  {journal} {\bibinfo  {journal} {Results in Physics}\ }\textbf
  {\bibinfo {volume} {45}},\ \bibinfo {pages} {106232} (\bibinfo {year}
  {2023})}\BibitemShut {NoStop}%
\bibitem [{\citenamefont {Ma}\ and\ \citenamefont {Susanto}(2021)}]{Ma2021}%
  \BibitemOpen
  \bibfield  {author} {\bibinfo {author} {\bibfnamefont {Y.-P.}\ \bibnamefont
  {Ma}}\ and\ \bibinfo {author} {\bibfnamefont {H.}~\bibnamefont {Susanto}},\
  }\bibfield  {title} {\bibinfo {title} {{Topological edge solitons and their
  stability in a nonlinear Su-Schrieffer-Heeger model}},\ }\href
  {https://doi.org/10.1103/PhysRevE.104.054206} {\bibfield  {journal} {\bibinfo
   {journal} {Physical Review E}\ }\textbf {\bibinfo {volume} {104}},\ \bibinfo
  {pages} {054206} (\bibinfo {year} {2021})}\BibitemShut {NoStop}%
\bibitem [{\citenamefont {Lugiato}\ and\ \citenamefont
  {Lefever}(1987)}]{Lugiato1987}%
  \BibitemOpen
  \bibfield  {author} {\bibinfo {author} {\bibfnamefont {L.~A.}\ \bibnamefont
  {Lugiato}}\ and\ \bibinfo {author} {\bibfnamefont {R.}~\bibnamefont
  {Lefever}},\ }\bibfield  {title} {\bibinfo {title} {{Spatial Dissipative
  Structures in Passive Optical Systems}},\ }\href
  {https://doi.org/10.1103/PhysRevLett.58.2209} {\bibfield  {journal} {\bibinfo
   {journal} {Physical Review Letters}\ }\textbf {\bibinfo {volume} {58}},\
  \bibinfo {pages} {2209} (\bibinfo {year} {1987})}\BibitemShut {NoStop}%
\bibitem [{\citenamefont {Haus}(1983)}]{Haus1983}%
  \BibitemOpen
  \bibfield  {author} {\bibinfo {author} {\bibfnamefont {H.~H.}\ \bibnamefont
  {Haus}},\ }\href@noop {} {\emph {\bibinfo {title} {Waves and Fields in
  Optoelectronics}}}\ (\bibinfo  {publisher} {Prentice Hall},\ \bibinfo {year}
  {1983})\BibitemShut {NoStop}%
\bibitem [{\citenamefont {Suh}\ \emph {et~al.}(2004)\citenamefont {Suh},
  \citenamefont {Wang},\ and\ \citenamefont {Fan}}]{Suh2004}%
  \BibitemOpen
  \bibfield  {author} {\bibinfo {author} {\bibfnamefont {W.}~\bibnamefont
  {Suh}}, \bibinfo {author} {\bibfnamefont {Z.}~\bibnamefont {Wang}},\ and\
  \bibinfo {author} {\bibfnamefont {S.}~\bibnamefont {Fan}},\ }\bibfield
  {title} {\bibinfo {title} {{Temporal coupled-mode theory and the presence of
  non-orthogonal modes in lossless multimode cavities}},\ }\href
  {https://doi.org/10.1109/JQE.2004.834773} {\bibfield  {journal} {\bibinfo
  {journal} {IEEE J. Quantum Electron.}\ }\textbf {\bibinfo {volume} {40}},\
  \bibinfo {pages} {1511} (\bibinfo {year} {2004})}\BibitemShut {NoStop}%
\bibitem [{\citenamefont {Agrawal}(2019)}]{Agrawal2019}%
  \BibitemOpen
  \bibfield  {author} {\bibinfo {author} {\bibfnamefont {G.}~\bibnamefont
  {Agrawal}},\ }\href@noop {} {\emph {\bibinfo {title} {{Nonlinear fiber
  optics}}}},\ \bibinfo {edition} {sixth}\ ed.\ (\bibinfo  {publisher}
  {Academic Press},\ \bibinfo {address} {London, England},\ \bibinfo {year}
  {2019})\BibitemShut {NoStop}%
\bibitem [{\citenamefont {Kaplan}\ and\ \citenamefont
  {Meystre}(1981)}]{Kaplan1981}%
  \BibitemOpen
  \bibfield  {author} {\bibinfo {author} {\bibfnamefont {A.~E.}\ \bibnamefont
  {Kaplan}}\ and\ \bibinfo {author} {\bibfnamefont {P.}~\bibnamefont
  {Meystre}},\ }\bibfield  {title} {\bibinfo {title} {{Enhancement of the
  Sagnac effect due to nonlinearly induced nonreciprocity}},\ }\href
  {https://doi.org/10.1364/OL.6.000590} {\bibfield  {journal} {\bibinfo
  {journal} {Optics Letters}\ }\textbf {\bibinfo {volume} {6}},\ \bibinfo
  {pages} {590} (\bibinfo {year} {1981})}\BibitemShut {NoStop}%
\bibitem [{\citenamefont {Cao}\ \emph {et~al.}(2020)\citenamefont {Cao},
  \citenamefont {Liu}, \citenamefont {Wang}, \citenamefont {Lu}, \citenamefont
  {Qiu}, \citenamefont {Rotter}, \citenamefont {Gong},\ and\ \citenamefont
  {Xiao}}]{Cao2020}%
  \BibitemOpen
  \bibfield  {author} {\bibinfo {author} {\bibfnamefont {Q.-T.}\ \bibnamefont
  {Cao}}, \bibinfo {author} {\bibfnamefont {R.}~\bibnamefont {Liu}}, \bibinfo
  {author} {\bibfnamefont {H.}~\bibnamefont {Wang}}, \bibinfo {author}
  {\bibfnamefont {Y.-K.}\ \bibnamefont {Lu}}, \bibinfo {author} {\bibfnamefont
  {C.-W.}\ \bibnamefont {Qiu}}, \bibinfo {author} {\bibfnamefont
  {S.}~\bibnamefont {Rotter}}, \bibinfo {author} {\bibfnamefont
  {Q.}~\bibnamefont {Gong}},\ and\ \bibinfo {author} {\bibfnamefont {Y.-F.}\
  \bibnamefont {Xiao}},\ }\bibfield  {title} {\bibinfo {title} {{Reconfigurable
  symmetry-broken laser in a symmetric microcavity}},\ }\href
  {https://doi.org/10.1038/s41467-020-14861-5} {\bibfield  {journal} {\bibinfo
  {journal} {Nature Communications}\ }\textbf {\bibinfo {volume} {11}},\
  \bibinfo {pages} {1136} (\bibinfo {year} {2020})}\BibitemShut {NoStop}%
\bibitem [{\citenamefont {Longhi}\ \emph {et~al.}(2018)\citenamefont {Longhi},
  \citenamefont {Kominis},\ and\ \citenamefont {Kovanis}}]{Longhi2018}%
  \BibitemOpen
  \bibfield  {author} {\bibinfo {author} {\bibfnamefont {S.}~\bibnamefont
  {Longhi}}, \bibinfo {author} {\bibfnamefont {Y.}~\bibnamefont {Kominis}},\
  and\ \bibinfo {author} {\bibfnamefont {V.}~\bibnamefont {Kovanis}},\
  }\bibfield  {title} {\bibinfo {title} {Presence of temporal dynamical
  instabilities in topological insulator lasers},\ }\href
  {https://doi.org/10.1209/0295-5075/122/14004} {\bibfield  {journal} {\bibinfo
   {journal} {EPL (Europhysics Letters)}\ }\textbf {\bibinfo {volume} {122}},\
  \bibinfo {pages} {14004} (\bibinfo {year} {2018})}\BibitemShut {NoStop}%
\bibitem [{\citenamefont {Bogaerts}\ \emph {et~al.}(2012)\citenamefont
  {Bogaerts}, \citenamefont {{De Heyn}}, \citenamefont {{Van Vaerenbergh}},
  \citenamefont {{De Vos}}, \citenamefont {{Kumar Selvaraja}}, \citenamefont
  {Claes}, \citenamefont {Dumon}, \citenamefont {Bienstman}, \citenamefont
  {{Van Thourhout}},\ and\ \citenamefont {Baets}}]{Bogaerts2012}%
  \BibitemOpen
  \bibfield  {author} {\bibinfo {author} {\bibfnamefont {W.}~\bibnamefont
  {Bogaerts}}, \bibinfo {author} {\bibfnamefont {P.}~\bibnamefont {{De Heyn}}},
  \bibinfo {author} {\bibfnamefont {T.}~\bibnamefont {{Van Vaerenbergh}}},
  \bibinfo {author} {\bibfnamefont {K.}~\bibnamefont {{De Vos}}}, \bibinfo
  {author} {\bibfnamefont {S.}~\bibnamefont {{Kumar Selvaraja}}}, \bibinfo
  {author} {\bibfnamefont {T.}~\bibnamefont {Claes}}, \bibinfo {author}
  {\bibfnamefont {P.}~\bibnamefont {Dumon}}, \bibinfo {author} {\bibfnamefont
  {P.}~\bibnamefont {Bienstman}}, \bibinfo {author} {\bibfnamefont
  {D.}~\bibnamefont {{Van Thourhout}}},\ and\ \bibinfo {author} {\bibfnamefont
  {R.}~\bibnamefont {Baets}},\ }\bibfield  {title} {\bibinfo {title} {{Silicon
  microring resonators}},\ }\href {https://doi.org/10.1002/lpor.201100017}
  {\bibfield  {journal} {\bibinfo  {journal} {Laser {\&} Photonics Reviews}\
  }\textbf {\bibinfo {volume} {6}},\ \bibinfo {pages} {47} (\bibinfo {year}
  {2012})}\BibitemShut {NoStop}%
\bibitem [{\citenamefont {Bino}\ \emph {et~al.}(2021)\citenamefont {Bino},
  \citenamefont {Moroney},\ and\ \citenamefont {Del'Haye}}]{DelBino:21}%
  \BibitemOpen
  \bibfield  {author} {\bibinfo {author} {\bibfnamefont {L.~D.}\ \bibnamefont
  {Bino}}, \bibinfo {author} {\bibfnamefont {N.}~\bibnamefont {Moroney}},\ and\
  \bibinfo {author} {\bibfnamefont {P.}~\bibnamefont {Del'Haye}},\ }\bibfield
  {title} {\bibinfo {title} {Optical memories and switching dynamics of
  counterpropagating light states in microresonators},\ }\href
  {https://doi.org/10.1364/OE.417951} {\bibfield  {journal} {\bibinfo
  {journal} {Opt. Express}\ }\textbf {\bibinfo {volume} {29}},\ \bibinfo
  {pages} {2193} (\bibinfo {year} {2021})}\BibitemShut {NoStop}%
\bibitem [{\citenamefont {Woodley}\ \emph {et~al.}(2021)\citenamefont
  {Woodley}, \citenamefont {Hill}, \citenamefont {{Del Bino}}, \citenamefont
  {Oppo},\ and\ \citenamefont {Del'Haye}}]{Woodley2020}%
  \BibitemOpen
  \bibfield  {author} {\bibinfo {author} {\bibfnamefont {M.~T.~M.}\
  \bibnamefont {Woodley}}, \bibinfo {author} {\bibfnamefont {L.}~\bibnamefont
  {Hill}}, \bibinfo {author} {\bibfnamefont {L.}~\bibnamefont {{Del Bino}}},
  \bibinfo {author} {\bibfnamefont {G.-L.}\ \bibnamefont {Oppo}},\ and\
  \bibinfo {author} {\bibfnamefont {P.}~\bibnamefont {Del'Haye}},\ }\bibfield
  {title} {\bibinfo {title} {{Self-Switching Kerr Oscillations of
  Counterpropagating Light in Microresonators}},\ }\href
  {https://doi.org/10.1103/PhysRevLett.126.043901} {\bibfield  {journal}
  {\bibinfo  {journal} {Physical Review Letters}\ }\textbf {\bibinfo {volume}
  {126}},\ \bibinfo {pages} {043901} (\bibinfo {year} {2021})}\BibitemShut
  {NoStop}%
\end{thebibliography}%
	
\end{document}